\documentclass[pdflatex,sn-mathphys-num]{sn-jnl}

\usepackage{todonotes}
\usepackage{bm}
\usepackage{graphicx}%
\usepackage{csquotes}
\usepackage{multirow}%
\usepackage{amsmath,amssymb,amsfonts}%
\usepackage{amsthm}%
\usepackage{mathrsfs}%
\usepackage[title]{appendix}%
\usepackage{xcolor}%
\usepackage{textcomp}%
\usepackage{manyfoot}%
\usepackage{booktabs}%
\usepackage{algorithm}%
\usepackage{algorithmicx}%
\usepackage{algpseudocode}%
\usepackage{listings}%
\usepackage{geometry, graphicx, fancyhdr}
\usepackage{booktabs}
\usepackage{longtable}
\usepackage{tikz}
\usepackage{pdflscape}
\usepackage{url}
\usepackage{caption}
\usepackage{mathtools}
\usepackage{subcaption}
\usetikzlibrary{arrows.meta, positioning, shapes.multipart}
\graphicspath{{./figs/}}

\theoremstyle{thmstyleone}%
\newtheorem{theorem}{Theorem}
\newtheorem{proposition}[theorem]{Proposition}%

\theoremstyle{thmstyletwo}%
\newtheorem{example}{Example}%
\newtheorem{remark}{Remark}%

\theoremstyle{thmstylethree}%
\newtheorem{definition}{Definition}%

\renewcommand{\b}[1]{\mathbf{#1}}
\begin{document}

\title{Using Machine Learning to Enhance Hyperparameter Optimization in Pandemic Modeling: Case study of COVID-19 Dynamics in Ghana}


\author*[1]{\fnm{Thomas} \sur{Izgin}}\email{izgin@mathematik.uni-kassel.de}

\author[1]{\fnm{Andreas} \sur{Meister}}\email{meister@mathematik.uni-kassel.de}

\author[2]{\fnm{Isaac} \sur{Azure}}\email{isaac.azure@knust.edu.gh}

\affil*[1]{\orgdiv{University of Kassel}, \orgname{Department of Mathematics and Natural Sciences}, \orgaddress{\street{Heinrich-Plett-Str. 40}, \city{Kassel}, \postcode{34132}, \state{Hesse}, \country{Germany}}}

\affil[2]{\orgdiv{Kwame Nkrumah University of Science and Technology}, \orgname{Department of Mathematics}, \orgaddress{\street{KNUST}, \city{Kumasi}, \postcode{AK-03220}, \state{Ashanti Region}, \country{Ghana}}}


\abstract{
  In this study, five distinct COVID-19 models developed in different countries, each designed to reflect the prevailing epidemiological condition at the time of formulation, are examined. The models are reformulated while still maintaining their original structure, using their common transmissions from one compartment to the other. Modified Patankar--Runge--Kutta (MPRK) methods are then applied to approximate the solutions of the resulting system of nonlinear ordinary differential equations (ODEs) representing each model to produce unconditionally positive approximations and to preserve the conservative part of the ODEs. In particular, we incorporate the numerical solution into a cost function to improve the estimates for the non-autonomous model hyperparameters. In a first step we obtain piecewise constant parameters that fit real data. Later we perform a WENO reconstruction in a post-process to approximate the true time-dependent coefficients inside the ODEs. As a proof-of-concept, we apply our approach to improve the parameters of a paper concerned with modeling COVID-19 in Ghana, where we can make 5-day predictions within a 10\% error range.
}




\maketitle

\section{Introduction}

COVID-19 is a new strain of the novel acute respiratory syndrome coronavirus $2$ (SARS-CoV-2) which is highly contagious and spreads faster through personal contact with infected individuals. As at January 21, 2020, four countries across the world had confirmed cases of COVID-19, with a total of $282$ cases and $6$ deaths. This data alerted the World Health Organization to officially declare COVID-19 as a pandemic; an event which ignited global research initiatives \cite{world2020novel}.

By March 26, 2020, close to 1.7 billion people worldwide were under some form of lockdown, and this number increased to 3.9 billion by the first week of April over half of the world’s population \cite{van2021cope}. According to \cite{kampf2020persistence}, COVID-19 aerosols, which serve as the main transmission medium, can remain active for up to 96 hours on surfaces. In contrast, other coronaviruses typically survive for about nine hours. The increasing numbers of COVID-19 cases makes it even more contagious and causes health and economic challenges to especially developing countries with less health facilities and equipments.

Consequently, several mathematical models have been developed globally to understand and predict the dynamics of COVID-19 transmission. These models aim to provide better strategies for disease control. Typically, such compartmental models divide the human population into groups such as Susceptible ($S$), Vaccinated $(V)$, Exposed ($E$), Latent ($L$) or Asymptomatic, Hospitalized ($H$), Infected ($I$) or Symptomatic, Recovered ($R$), Quarantine $(Q)$ and Virus in the Environment ($V_E$). The $S$ population represents individuals who have not yet been infected and are not vaccinated, but can contract the virus upon exposure to an infected person not vaccinated, $V$ represents the population that has been vaccinated and is not currently exposed, the $E$ population are those who are in contact with individuals who may have contracted the virus, the $L$ population are those who the disease incubates, and hence are those who are infected but do not exhibit clinical or noticeable symptoms, although they can transmit the virus to others and into the environment. The $I$ population is those identified as being infected by the virus but not in $H$. The $H$ population are those who have been diagnosed with COVID-19 and are currently receiving treatment at the hospital, the $Q$ population are those who have been isolated for recovery. The $R$ population are the people who have recovered from the disease and the $D$ compartment represents those who died as a result natural death while $D_\text{covid}$ accounts for COVID-19 disease fatalities, and the $V_E$ compartment represents the virus emitted by the virus carriers into the environment.

The study \cite{Ramalingam2025Stability} improved the traditional SEIR COVID-19 model by incorporating additional epidemiological factors such as detection, diagnosis, quarantine measures, vaccination, and other pharmaceutical and non-pharmaceutical interventions to better capture the dynamics of the pandemic. 
The model was further enhanced through rigorous equilibrium and stability analyses based on the basic reproduction number, as well as sensitivity analysis to determine the influence of transmission parameters on disease spread. In addition, the researchers integrated optimal control theory using Pontryagin’s Maximum Principle to minimize infections, fatalities, and intervention costs simultaneously. The combination of deterministic mathematical analysis with numerical simulations increased the reliability, applicability, and predictive capability of the model, enabling a more comprehensive understanding of COVID-19 transmission dynamics and supporting effective public health decision-making and resource allocation. In the related study  \cite{Li2024Extended}, the traditional SEIR epidemic model was improved by developing an extended SEIR framework with eight state variables that captures the movement of individuals across households, communities, and hospitals during the spread of COVID-19. Unlike earlier SEIR models that assumed a homogeneous spatial distribution of the population, the proposed model incorporated social-domain interactions, providing a more realistic representation of disease transmission dynamics. The model was further strengthened through the analysis of important mathematical properties such as the basic reproduction number, disease-free equilibrium, endemic equilibrium, and their local stability. In addition, the researchers calibrated the model parameters using the Markov Chain Monte Carlo (MCMC) method, which enhanced the accuracy and reliability of the model in reproducing the spread of the Delta variant outbreak in England. Sensitivity analysis was also employed to identify effective control measures, making the model useful for guiding public health management and COVID-19 intervention strategies.

In recent times, the introduction of machine learning has improved the predictability of mathematical models for COVID-19 related researches. A study by \cite{Alkhalefah2024Deep}, developed a hybrid modeling framework that integrated a non-autonomous Susceptible–Infected–Recovered–Vaccinated–Deceased (SIRVD) epidemiological model with machine learning and deep learning techniques. However, the time-dependent coefficients were estimated by a single forward Euler step with a time step size of one day, which may result in negative estimates.  
In the study by \cite{Qian2025Physics-informed}, infectious disease forecasting was improved by implementing a physics-informed neural network (PINN) framework that integrated epidemiological compartmental modeling with deep learning. Unlike standard data-driven models, the PINN incorporated the dynamical system equations of disease transmission directly into the loss function.  While this approach takes time as an input and generates approximations to the solution of the ODEs as well as estimates for the model parameters, this technique does not give an estimate on the error of the approximation for the underlying model. In particular, conservation of the total population count or positivity of the modeled quantities may be violated in theory.
For further insights and works in this field, we refer to \cite{Shivalingappa2023Prediction,Alizadeh2025Epidemic,Adewole2024COVID-19}, the review article \cite{Cheng2025Machine} and the literature mentioned therein.

This study consists of two parts. First of all, we present a comprehensive model generalizing generalizing the before-mentioned ODE, and give the parameters to reconstruct the models mentioned in \cite{moore2022global,diagne2021mathematical,santosh2025novel,rattanakul2024mathematical,haq2022new}.  These included a model for Ghana \cite{moore2022global}, comprising six compartments ($S, E, L, I, R, V_E$). A seven-compartment model designed for Senegal included 
$(S, V, E, I, L, H, R)$, \cite{diagne2021mathematical}. Furthermore, two other models formulated for the United States with the compartments $(S, V, E, I, H, R)$ and $(S, E, I, H, R)$ respectively \cite{santosh2025novel,rattanakul2024mathematical}, and a fifth model with compartments $(S,V,E,I,Q,R)$ designed for Pakistan \cite{haq2022new}. 
Each of these models inherently captured a key physical property, that is, the non-negativity of the involved compartments. It inherits also a conservative part as people are mostly transferred from one compartment to the other. However, the models are not conservative unless recruitment and deaths are captured properly and $V_E$ is not part of the conservativity requirement. In an abstract setting, the related systems of differential equations may be interpreted as a so-called production-destruction-rest system (PDRS), which highlights the conservative and nonconservative parts, enabling tailored numerical treatment to the respective parts. 

The second part constitutes the main contribution of this work and consists of developing a code \cite{AIM2026repository} and technique that has the potential to overcome the issue of works that incorporate the ODEs inside the cost function rather than its numerical approximation. In particular, rather than using the ODE solely as a penalty term when fitting curves to raw data, we fit the ODE parameters, and thus, the ODEs itself to fit the raw data to make it a good model for the pandemic or epidemic situation. We then then solve the resulting ODEs numerically, which enables us to harness the rich theory for numerical methods and their properties. In this regard, we use Modified Patankar Runge-Kutta (MPRK) schemes \cite{BDM2003,kopecz2018order, izgin2024unifying} as a time integrator, which are designed to approximate the solution of a PDRS while guaranteeing the positivity of the solution and preserving the conservative part of the PDRS. The latter may be understood as \emph{local conservation}.  To be more precise, we develop a cost function that takes the hyperparameters as an input and computes a numerical solution for the system of ODEs, which is then compared with real data, yielding the respective costs. In lack of a derivative of this complex cost function, we use Bayesian optimization to find a minimum. 

In particular, the work is organized as follows. Section~\ref{sec:model_form} introduces the formulation of the model and recasts it as PDRS. Section~\ref{sec:MPRK} presents the numerical method, which fits the needs of the ODEs. Section~\ref{sec:cost} discusses the cost function and provides notes on the implementation, which is available at \cite{AIM2026repository}. Section~\ref{sec:num_exp} presents the numerical results, while Section~\ref{sec:summary} concludes the work with a summary and outlook for future research.

\section{Model Formulation}\label{sec:model_form}
To gain a deeper understanding of the dynamics underlying COVID-19 transmission, this study developed a comprehensive model that integrates and extends at least five existing models \cite{moore2022global,diagne2021mathematical,santosh2025novel,rattanakul2024mathematical,haq2022new}. The general model incorporates ten key compartments $S, V, E, L, I, H, R, Q, D, D_\text{covid}$ and $V_E$ to capture all essential processes relevant to the control and spread of the epidemic. By doing so, each of the five individual models could be represented as a special case of this generalized system of nonlinear differential equations. Figure~\ref{fig:covid19model} below presents the developed comprehensive COVID-19 model, illustrating all state variables and parameters.

The model employs several parameters to describe population transition and disease progression. Specifically, $\delta_V, \delta_E, \delta_L, \delta_H,$ and $\delta_Q$ denote parameters that switch on or off the $V, E, L, H,$ and $Q$ compartments, respectively. The parameter $p$ represents the proportion of newly recruited individuals who are vaccinated, while $\mu$ is the natural death rate. The parameters $\phi$ and $\gamma$ describe the fractions of exposed and latent individuals progressing to infection, and $\psi$ denotes the fraction of infected individuals who recover without hospitalization. The rate $\mu_{V_E}$ corresponds to the decay of the virus in the environment, whereas $\alpha_I$ and $\alpha_H$ represent the disease-induced death rates for the infected and hospitalized populations, respectively. Finally, $a_S^V$ and $a_V^S$ describe transitions from  $V$ to $S$ and from $S$ to $V$ compartments respectively, with similar interpretations applying to the other parameters bearing subscripts and superscripts. Table~\ref{tab:parameters} describes all the parameter notations used in the model, while Figure~\ref{fig:covid19model} shows the outlook of the generalized COVID-19 model out of which the model equation \eqref{eq:compr_model} is obtained. The respective models from \cite{moore2022global,diagne2021mathematical,santosh2025novel,rattanakul2024mathematical,haq2022new} are recovered by using the specific parameters from Table~\ref{tab:model_para} in the Appendix~\ref{app:model_para}.



\begin{figure}[!htbp]
	\centering
\includegraphics[width=\textwidth, height=0.9\textheight, keepaspectratio]{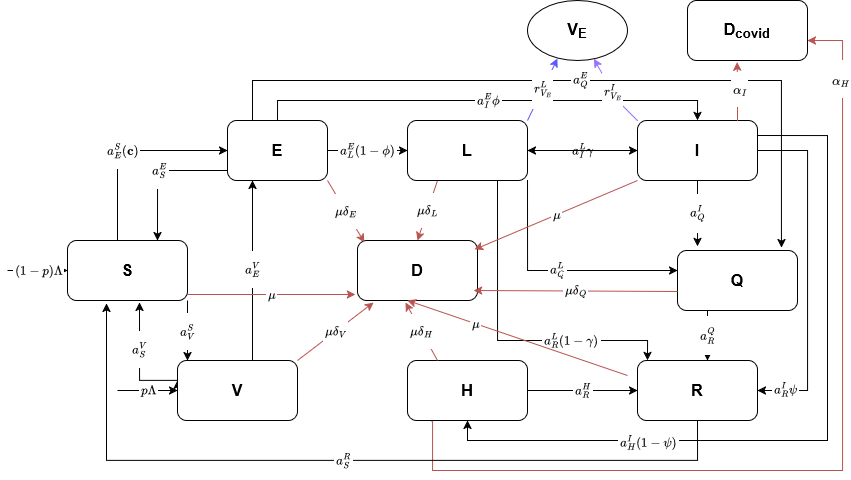}
	\caption{Generalized COVID-19 model for the selected models \cite{moore2022global,diagne2021mathematical,santosh2025novel,rattanakul2024mathematical,haq2022new}.}
	\label{fig:covid19model}
\end{figure}

\begin{equation}\label{eq:compr_model}
	\begin{aligned}
		\frac{dS}{dt} &= (1-p)\Lambda + a_S^V V  +a_S^E E - a_E^S(\b u)S - \mu S - a_V^S S + a_S^R R, \\[1em]
		\frac{dV}{dt} &= p\Lambda + a_V^S S - a_S^V V - a_E^V(\b u) V - \mu\delta_V V, \\[1em]
		\frac{dE}{dt} &= a_E^S(\b u)S + a_E^V(\b u) V - a_I^E \phi E -a_S^E E- a_Q^E E - a_L^E (1-\phi)E - \mu\delta_E E, \\[1em]
		\frac{dL}{dt} &= a_L^E (1-\phi)E - a_I^L \gamma L - a_R^L (1-\gamma)L - a_Q^L L - \mu\delta_L L, \\[1em]
		\frac{dI}{dt} &= a_I^E \phi E + a_I^L \gamma L - a_H^I (1-\psi)I - a_R^I \psi I - a_Q^I I - (\mu+\alpha_I)I, \\[1em]
		\frac{dH}{dt} &= a_H^I (1-\psi)I - a_R^H H - (\mu\delta_H+\alpha_H)H, \\[1em]
		\frac{dR}{dt} &= a_R^L (1-\gamma)L + a_R^I \psi I + a_R^Q Q + a_R^H H - a_S^R R - \mu R , \\[1em]
		\frac{dQ}{dt} &= a_Q^E E + a_Q^L L + a_Q^I I - a_R^Q Q - \mu\delta_Q Q, \\[1em]
		\frac{dD}{dt} &= \mu\left( S + \delta_E E + \delta_L L +I + \delta_H H + R + \delta_Q Q + \delta_V V\right) , \\[1em]
                \frac{dD_\text{covid}}{dt} &=\alpha_I I +\alpha_H  H,\\[1em]
		\frac{dV_E}{dt} &= r_{V_E}^L L + r_{V_E}^I I - \mu_{V_E} V_E,
	\end{aligned}
\end{equation}
where the coefficients other than $\delta_X$ with $X\in\{V,E,L,H,Q\}$ may be time-dependent. However, we omitted to write the time dependency for a better presentation and reading flow.
Here and in the following, $\b u$ is the state vector, that is
\[
\mathbf{u} = 
\left(
u_1, u_2, u_3, u_4, ..., u_{11}
\right)^T
= 
\left(
S, V, E, L,..., V_E
\right)^{T}.
\]
We point out that, according to Table~\ref{tab:model_para} in the appendix, the solution-dependency of $a^S_E$ and $a^V_E$ can be written as
\[a^S_E(\b u)= (\b z^T \b u) \cdot (\b u^T\b 1-D- D_\text{covid}- V_E)^{k_1},\quad  a^V_E(\b u) = z^V_E (a^S_E(\b u))^{k_2},\]
where $k_1\in\{-1,0,1\}$, $k_2\in \{0,1\}$, $\b 1=(1,\dotsc,1)^T$ and $\b z, z^V_E$ are time-dependent coefficients.
To outline our idea, we first interpret the system \eqref{eq:compr_model} as a so-called production-destruction-rest system.

\subsection{Production-Destruction-Rest System for the Comprehensive Model}
According to \cite{IR2023}, every system of ODEs with real-valued right-hand sides can be recast as a so-called production-destruction-rest system (PDRS),
\begin{equation}
		u_i'(t)= r_i(\b u(t), t) + \sum_{j=1}^N (p_{ij}(\b u(t), t)-d_{ij}(\b u(t), t)),\quad \b u(0)=\b u^0\in \mathbb R^N_{>0}, \label{eq:PDRS}
	\end{equation}
where $p_{ij}(\b u(t),t), d_{ij}(\b u(t),t) \geq 0$ for all $\b u(t)\geq\b 0$, $t\geq 0$ for $i,j=1,\dotsc,N$. Here and in the following, vector inequalities as well as operations on vectors such as summation, division and squaring should be understood in a component-wise manner. The expressions $p_{ij}$ and $d_{ij}$ represent production and destruction terms, respectively, while $r_i$ refer to rest terms. In this notation, we require that the PDS part obtained by neglecting the rest terms is \textit{conservative}, i.e., it satisfies $p_{ij}=d_{ji}$ as well as $p_{ii}=d_{ii}=0$ for $i,j=1,\dotsc,N$. Following \cite{IMST2026}, the rest terms are also split for $i=1,\dotsc,N$ according to
\begin{equation}
    r_i(\b u(t), t) = r_i^p(\b u(t), t) - r_i^d(\b u(t), t)\label{eq:rp_rd}
\end{equation}
with $r_i^p,r_i^d\geq 0$ for $t\geq 0$ and  $\b u(t)\geq \b 0$. This reformulation of the ODEs allow a numerical method to control the conservative part while guaranteeing positivity, as we will see in Section~\ref{sec:MPRK}.

Regarding the comprehensive model represented by equation \eqref{eq:compr_model}, the nonzero production terms are
\[
\begin{aligned}
	p_{12} &= a_S^V V, \quad &
	p_{13} &= a_S^E E, \quad &
	p_{17} &= a_S^R R, \\[0.5em]
	p_{21} &= a_V^S S, \quad &
	p_{31} &= a_E^S (\b{u})S, \quad &
	p_{32} &= a_E^V(\b u) V, \\[0.5em]
	p_{43} &= a_L^E (1-\phi)E, \quad &
	p_{53} &= a_I^E \phi E, \quad &
	p_{54} &= a_I^L \gamma L, \\[0.5em]
	p_{65} &= a_H^I (1-\psi)I, \quad &
	p_{74} &= a_R^L (1-\gamma)L, \quad &
	p_{75} &= a_R^I \psi I, \\[0.5em]
	p_{76} &= a_R^H H, \quad &
	p_{78} &= a_R^Q Q, \quad &
	p_{83} &= a_Q^E E, \\[0.5em]
	p_{84} &= a_Q^L L, \quad &
	p_{85} &= a_Q^I I, \quad &
	p_{91} &= \mu S, \\[0.5em]
	p_{92} &= \mu \delta_V V, \quad &
	p_{93} &= \mu \delta_E E, \quad &
	p_{94} &= \mu \delta_L L, \\[0.5em]
	p_{95} &= (\mu + \alpha_I)I, \quad &
	p_{96} &= \mu \delta_H , \quad &
	p_{97} &= \mu R, \\[0.5em]
	p_{98} &= \mu \delta_Q Q, \quad & p_{10,5}&= \alpha_I I,\quad & p_{10,6}&= \alpha_H H,
\end{aligned}
\]

The rest terms for the comprehensive model represented by equation \eqref{eq:compr_model} are

\[
r_1^p = (1-p)\Lambda, \quad 
r_2^p = p\Lambda, \quad 
r_{11}^p = r_{V_E}^L L + r_{V_E}^I I, \quad r_{11}^d = \mu_{V_E} V_E.
\]

\section{Modified Patankar--Runge-Kutta Schemes}\label{sec:MPRK}
Unlike the general linear methods of higher order, the Modified Patankar-Runge-Kutta schemes (MPRK) are unconditionally positivity-preserving methods, and since equation \eqref{eq:compr_model} has non-negative state variables, MPRK schemes are suitable to numerically solve this model. According to \cite{izgin2024unifying}, the general MPRK scheme for any system of ordinary differential equations written in a form of production--destruction-rest system is based on a single explicit Butcher array determined by $\b A, \b b,\b c\geq \b 0$  and reads

\begin{equation}
	u_i^{(k)} = u_i^n
	+ \Delta t \sum_{\nu=1}^{k-1} a_{k\nu} 
	\left(r_i^p(\b{u}^{(\nu)},t_k) +
	\sum_{j=1}^{N} p_{ij}(\b{u}^{(\nu)},t_k) \frac{u_j^{(k)}}{\pi_j^{(k)}}
	- 
	\left( r_i^D(\b{u}^{(\nu)},t_k) + \sum_{j=1}^{N} d_{ij}(\b{u}^{(\nu)},t_k) \right) 
	\frac{u_i^{(k)}}{\pi_i^{(k)}}
	\right)
	,
\end{equation}
for $k = 1, \dotsc, s,$ $t_k=t_n + c_k\Delta t$, and the final step is computed by
\begin{equation}
	u_i^{n+1} = u_i^n 
	+ \Delta t \sum_{k=1}^{s} b_k 
	\left(r_i^p(\b{u}^{(k)},t_k) +
	\sum_{j=1}^{N} p_{ij}(\b{u}^{(k)},t_k) \frac{u_j^{n+1}}{\sigma_j}
	- 
	\left( r_i^D(\b{u}^{(k)},t_k) + \sum_{j=1}^{N} d_{ij}(\b{u}^{(k)},t_k) \right) 
	\frac{u_i^{n+1}}{\sigma_i}
	\right).
\end{equation}

There, the so-called Patankar weight-denominators (PWDs) $\bm\pi^{(k)}=\bm \pi^{(k)}(\b u^n, \b u^{(1)},\dotsc,\b u^{(k-1)})$ and $\bm \sigma=\bm \sigma(\b{u}^n, \b{u}^{(1)}, \ldots, \b{u}^{(s)})$ are free parameters characterized by

\[
\pi_i^{(k)}, \sigma_i > 0 
\quad \text{for } \b u^n > 0
\]
when interpreted as a function of $\b u^n$.
There is extensive research on how to choose them properly to obtain higher accuracy \cite{kopecz2018order, MPDeC,NSARK, izgin2024unifying} or better stability properties \cite{IKM2022b,IOE22StabMP,izgin2024unifying} to name just a few.


\noindent
\begin{example}[Modified Patankar Euler]
    Choosing the explicit Euler method as the underlying scheme, we have $s=1$. Note, that in this case, $\pi_i^{(1)}$ has no effect. Then, choosing $\sigma_j = u_j^n$, yields the first-order accurate MPRK scheme which is also called the Modified Patankar-Euler scheme (MPE) \cite{BDM2003}, and reads

\begin{equation}\label{eq:MPE}
	u_i^{n+1} = u_i^n + \Delta t\, r_i^p(\b{u}^n,t_n)
	+ \Delta t \sum_{j=1}^{N} p_{ij}(\b{u}^n,t_n) \frac{u_j^{n+1}}{u_j^n}
	-\Delta t \left( r_i^D(\b{u}^n,t_n) + \sum_{j=1}^{N} d_{ij}(\b{u}^n,t_n) \right) 
	\frac{u_i^{n+1}}{u_i^n}, \quad i=1,\dotsc,N.
\end{equation}

\end{example}
\begin{example}[Modified Patankar Heun]
    Although there is a one-parameter family of second-order MPRK schemes \cite{kopecz2018order}, we follow the suggestions from \cite{IssuesMPRK} and we will use the scheme based on Heun's method using the Patankar weights $\bm \pi^{(2)}=\b u^n$ and $\bm \sigma=\b u^{(2)}$. The corresponding scheme is denoted by MPRK22(1), see \cite{kopecz2018order} for more details.
\end{example}

\begin{remark}[Matrix Representation and Positivity]
The MPE scheme form in equation \eqref{eq:MPE} can be written as
\begin{equation}
	\b M(\b u^n) \b u^{n+1}
	= 
	u_{i}^{n} + \Delta t\, r_{i}^{p}\left(\b{u}^{n},t_n\right),
\end{equation}
where the entries $m_{ij}$ of the solution-dependent matrix $\b M$ are given by
\begin{equation}
	m_{ii} = 1 + 
	\frac{\Delta t \sum_{j=1}^{N} d_{ij}\left(\b{u}^{n}t_n\right) + r_{i}^{D}\left(\b{u}^{n},t_n\right)}{u_{i}^{n}},
	\qquad
	m_{ij} = -\frac{\Delta t\, p_{ij}\left(\b{u}^{n},t_n\right)}{u_{j}^{n}}, \quad i\neq j.
\end{equation}
It is proven in \cite{BDM2003} that the mass matrix $\b M$ is an M-matrix, that is, it possesses a non-negative inverse for all $\Delta t>0$. This key property yields the unconditional positivity of the scheme. Indeed, all MPRK schemes are structurally the same and it is proven by induction and along the same lines that even the mass matrices for the stage vectors $\b u^{(k)}$ are M-matrices, see \cite{kopecz2018order, izgin2024unifying}.
\end{remark}

\section{Prediction Procedure}\label{sec:cost} 
In this study, we focus on Ghana, and as a proof-of-concept, on the model from \cite{moore2022global} with the compartments $S, E, L, I, R, V_E$. Out of these, the number of infected people, $I$, are measurable. As our comprehensive model additionally accounts for dead people in compartment $D$, we can also incorporate data for this compartment. To be precise, we introduce a new compartment $D_\text{covid}$ for our implementation to keep track of the fatalities caused by the disease.  The data for these compartments are taken from \cite{data_D_Ghana,data_I_Ghana}\footnote{At the time of access, the data in \cite{data_D_Ghana} has a flaw depicting 0 confirmed deaths until September 25, 2020, but reporting 301 deaths the day after. In contrast, \cite{data_I_Ghana} also provides data prior to this date.} and incorporated into our Julia code \cite{AIM2026repository}.
\subsection{Estimation Algorithm for Coefficients of the Model and Cost Function}
As can be seen in Table~\ref{tab:model_para}, some of the coefficients are zero as not all compartments play a role. Also, we assume that the recruitment rate $\Lambda$ and natural death rate $\mu$ are well estimated and constant throughout a year, leaving 15 time-dependent parameters unknown. These include the coefficients $c_I, c_L, c_{V_E}$ of the respective compartments in $a^S_E$. Together they form the parameter vector $\b p(t)$. 

As we do not know the time evolution of $\b p$, we estimate it by first subdividing the time span $I=[0, T]$ for some $T>0$ into intervals $I_1,\dotsc, I_M$ such that $I=\cup_{i=1}^M I_i$, $ I^\circ_i\cap I^\circ_j=\emptyset$ and computing \textit{optimal} constant coefficients for each $I_i$. We do that by minimizing a cost function that takes a constant parameter vector $\bar{\b p}$ as an input. To that end, we first compute a numerical solution of the respective system of ODE substituting $\b p(t)$ by the constant $\bar{\b p}$, and comparing the numerical solution with real data. In particular, we compute a $\b u^n$ using an adaptive time stepping with MPRK22(1), and apply linear interpolation to recover second-order and positive approximations $\b u^r$ at those points in time, where we have a real data $\b u^\text{data}$. Note that this procedure can be done using higher order MPRK schemes, too, by replacing the linear interpolation by a positivity-preserving dense output formula \cite{Izgin2026BootStrapping}. We  then extract the data for the compartments $I$ and $D_\text{covid}$ at the respective times forming new vectors, which we denote by $\b I^\text{num}$, $\b D_\text{covid}^\text{num}$ and $\b I^\text{data}$, $\b D_\text{covid}^\text{data}$ and compute the costs $c(\bar{\b p})$ by
\[c(\bar{\b p}) = \frac{\Vert \b I^\text{num}- \b I^\text{data}\Vert}{\Vert\max(\b I^\text{data},1)\Vert}+ \frac{\Vert \b D_\text{covid}^\text{num}- \b D_\text{covid}^\text{data}\Vert}{\Vert\max(\b D_\text{covid}^\text{data},1)\Vert},\]
where $\max(\b x,1)=(\max(x_i,1),\dotsc,\max(x_M,1))^T$ for $\b x\in \mathbb R^M$. We note that the real data are in $\mathbb N_0$, so that the cost value in most cases reflects the relative error. The minimum of this cost function is searched via Bayesian optimization \cite{Bull2011,GSA2014, SLA2012}, where we use two different acquisition functions related to \textit{expected-improvement} and \textit{probability-of-improvement}, for a good trade-off of exploration and local exploitation. This optimization tool only evaluates the cost function, and after 150 iterations with each acquisition function, the algorithm begins a new cycle if the value of $c$ is improved by $10^{-3}$. We start the optimization with an time interval of 10 days and check if $c(\bar{\b p})<0.25$. If not, the time interval shrinks to 5 days, then 3 days. Finally, our algorithm takes the interval with the lowest cost, stores the parameter vector and continues with the next time interval.
\subsubsection{Reconstruction of the Time-Dependent Coefficients}
After computing the best piecewise constant parameter vector over the intervals $I_1,\dotsc,I_M$ with $I_l=[t_{l-\frac12},t_{l+\frac12}]$, $\Delta t_l=\lvert I_l\rvert=t_{l+\frac12}-t_{l-\frac12}$ for $l=1,\dotsc,M$, we interpret the piecewise constant values as cell averages and reconstruct the time-dependent approximation to $\b p(t)$ by means of a WENO reconstruction \cite{LiuOsherChan1994,JiangShu1996}.

In particular, we implemented the following for $k\in \{2,3\}$, which corresponds to the WENO method of order $2k-1$. Given a cell $I_i$, we consider $r=0,\dotsc,k-1$ together with the stencils
\begin{equation*}
S_r(i) = \{I_{i-r}, I_{i-r+1}, \dots, I_{i+k-r-1}\}
\end{equation*}
and the respective set of cell averages 
\[ \{\bar{\b p}_{i-r},\dotsc, \bar{\b p}_{i+k-r-1}\}.\]
We form the polynomial
\begin{equation}\label{eq:p_tilde}
    \widetilde{\b p}^i_r(t)= \sum_{m=0}^k\frac{\sum_{\substack{l=0\\l\neq m}}^k\prod_{\substack{q=0\\q\neq m,l}}^k(t-t_{i-r+q-\frac12})}{\prod_{\substack{l=0\\l\neq m}}^k(t_{i-r-m-\frac12}-t_{i-r+l-\frac12})}\sum_{j=0}^{m-1} \bar{\b p}_{i-r+j} \Delta t_{i-r+j}.
\end{equation}
and compute the unique (vector) polynomial  $\b P^i$ of degree $2k-1$ satisfying 
\[\frac{1}{\Delta t_l}\int_{I_l} \b P^i(\tau)\mathrm d\tau= \bar{\b p}^i_l,\quad l=i-k+1,\dotsc, i+k-1,\]
which can also be found using \eqref{eq:p_tilde} with $r=k-1$ and extending the upper limits from $k$ to $2k-1$.
Then we obtain the functions $\b d^i_r(t)$ satisfying
\[\sum_{r=0}^{k-1}\b d^i_r(t)=\b 1,\quad \b P^i(t)=\sum_{r=0}^{k-1}\b d^i_r(t)\widetilde{\b p}^i_r(t)\] by solving the corresponding augmented linear system for any given $t$. We continue computing the vector of smoothness indicators 
\begin{equation*}
    \bm\beta^i_r = \sum_{l=1}^{k-1} \int_{I_i} \Delta t_i^{2l-1} \left( \frac{\partial^l \widetilde{\b p}^i_r(\tau)}{\partial x^l} \right)^2 \, \mathrm d\tau.
\end{equation*}
Finally, the WENO approximation for $t\in I_i$ is given by
\begin{equation*}
    \b p^i_{\text{WENO}}(t) = \sum_{r=0}^{k-1} \bm\omega^i_r(t) \widetilde{\b p}^i_r(t),
\end{equation*}
where
\begin{equation*}
   \bm\omega^i_r(t) = \frac{\bm\alpha^i_r(t)}{\sum_{m=0}^{k-1} \bm\alpha^i_m(t)},\quad  \bm\alpha^i_r(t) = \frac{\bm d^i_r(t)}{(\bm\beta^i_r + \varepsilon)^2}, \quad \epsilon=10^{-6}, \quad r=0,\dotsc,k-1.
\end{equation*}
We emphasize again that the operations involving vectors are executed component-wise.
To preserve positivity as well as an upper bound, we also implemented the Zhang--Shu limiter from \cite{ZhangShu2010MPS}. To make sure this procedure is well-defined at the boundaries, we introduce $k-1$ ghost cells $I_{-(k-2)},\dotsc,I_0, I_{M+1},\dotsc,I_{M+k-1}$ by means of constant or linear extrapolation of the cell averages. The boundary reconstruction $\b p^M_{\textrm{WENO}}$ is used for extrapolation, if $t>t_{M+\frac12}$. However, we note that the accuracy of the reconstruction drops, if it is evaluated at $t>t_{M+k-\frac12}$.
\section{Numerical Simulation}\label{sec:num_exp}
\begin{figure}[!htbp]
    \centering
\includegraphics[width=0.5\linewidth]{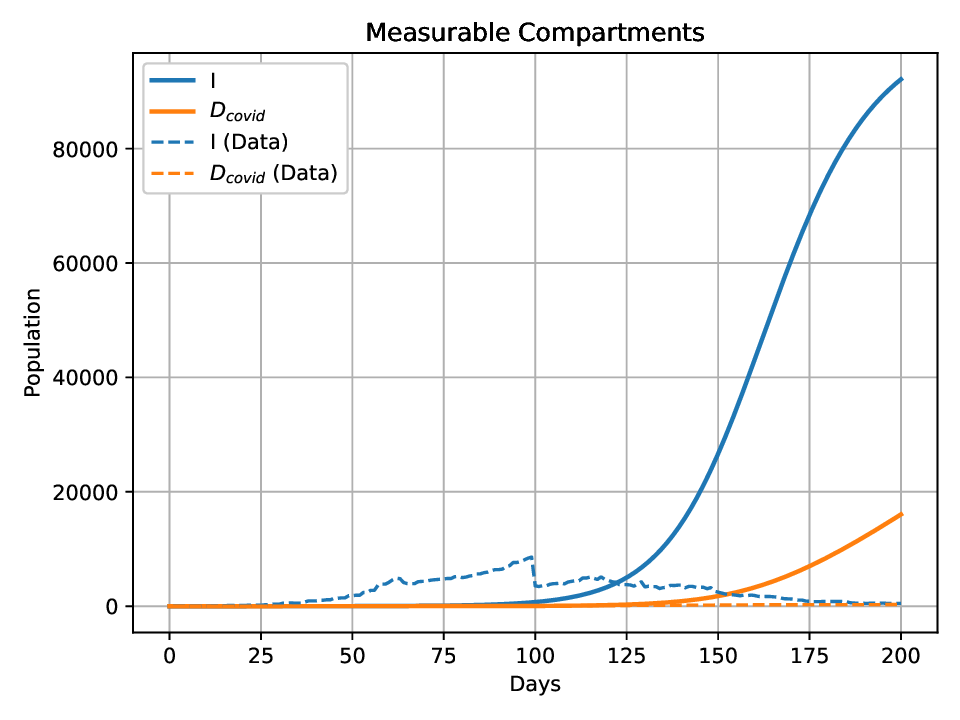}
    \caption{Solution of \eqref{eq:compr_model} with constant Model~1 parameters from Table~\ref{tab:model_para} and initial condition from Table~\ref{tab:IC} \cite{moore2022global}.}
    \label{fig:model1}
\end{figure}

\begin{figure}[!htbp]
    \centering
    \includegraphics[width=\linewidth]{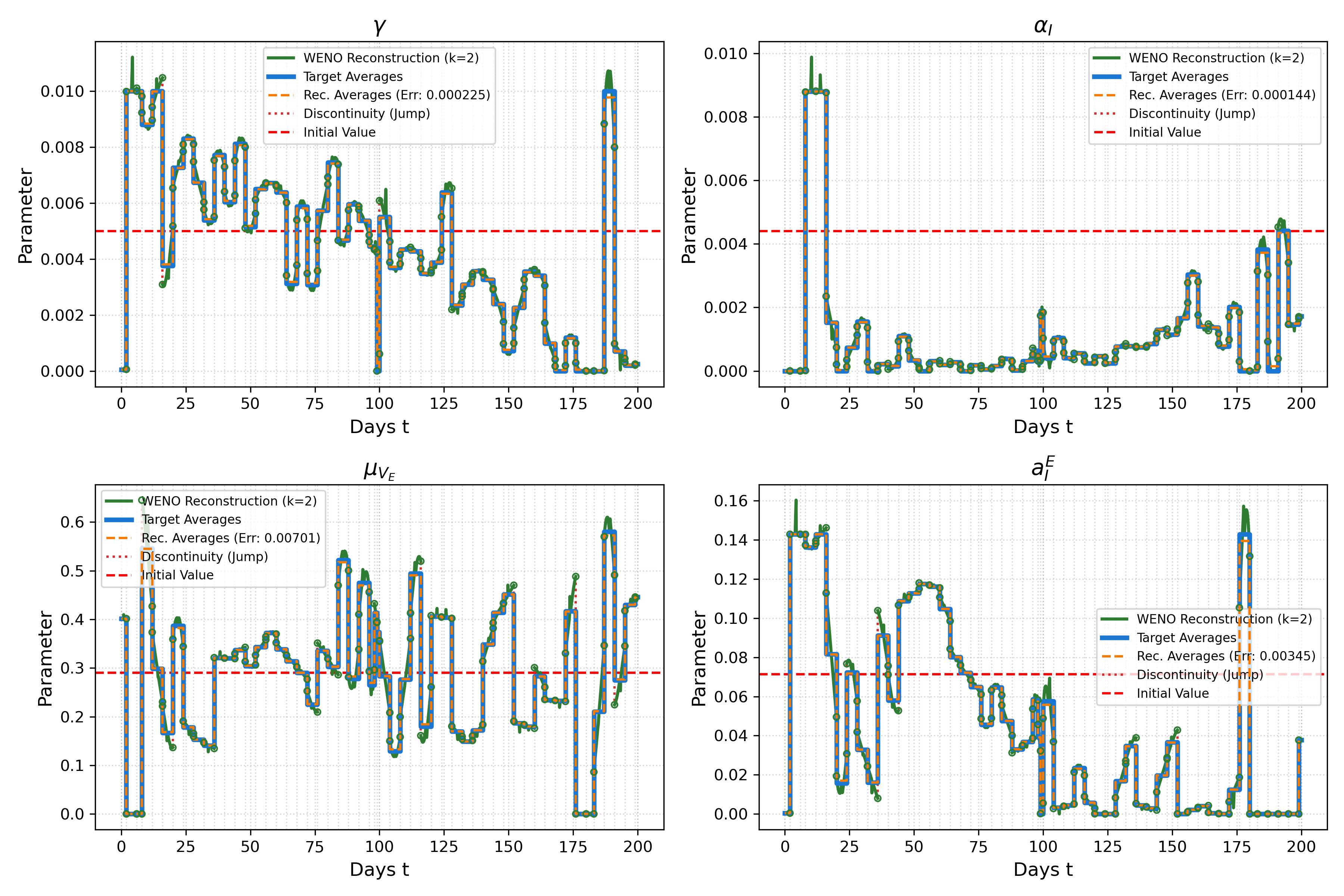}
    \caption{Reconstructed parameters using the third-order WENO procedure. The initial values are from Table~\ref{tab:model_para} \cite{moore2022global}.}
    \label{fig:weno_para}
\end{figure}

\begin{figure}[!htbp]
    \centering
    \includegraphics[width=0.75\linewidth]{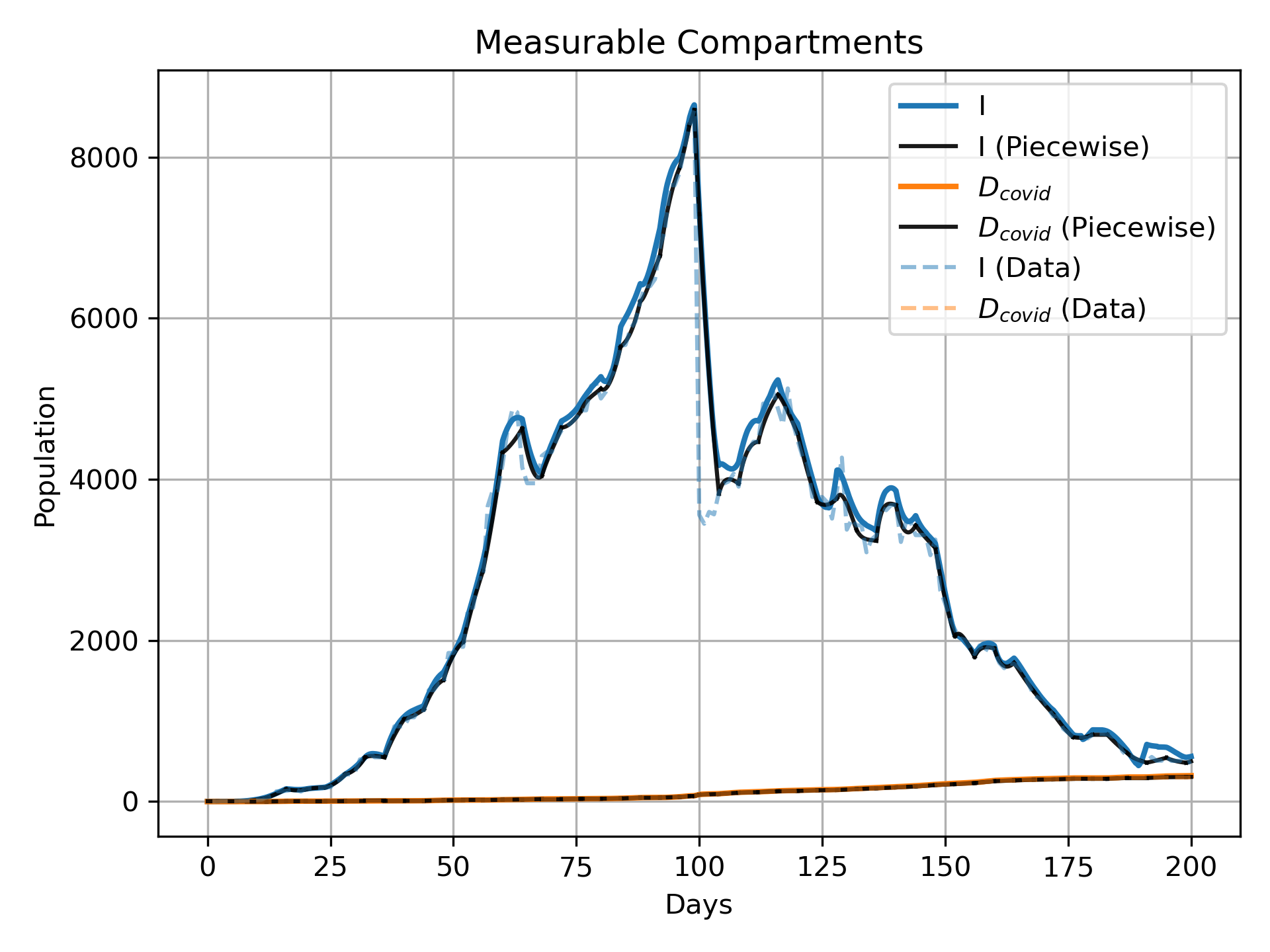}
    \caption{Solution of \eqref{eq:compr_model} with improved Model~1 parameters as well as their WENO reconstruction.}
    \label{fig:PW_and_WENO_sol}
\end{figure}

As mentioned before, we considered the model problem from \cite{moore2022global}, which assumed constant coefficients. Consequently, it fails dramatically to fit the real data, see Figure~\ref{fig:model1}. Here, the corresponding initial condition can be read off from Table~\ref{tab:IC}.
In contrast, we can find piecewise constant parameters using our novel cost function and then reconstruct time-dependent parameters by the procedure outlined in the previous section. Figure~\ref{fig:weno_para} gives illustrates the results for some of the parameters of Table~\ref{tab:model_para}.
There, the recovered averages are computed using a quadrature rule, which is why the legend does not show an error of 0. Instead, the maximum error for all the cells is given. Now, we are in the position to solve the non-autonomous system of ODEs using piecewise constant as well as the WENO reconstructed parameters, see Figure~\ref{fig:PW_and_WENO_sol}.
\subsection{Predictability of Future COVID-19 Dynamics}
We consider the following techniques to make a forecast of the pandemic dynamics in Ghana for $L$ days with corresponding time intervals $I_{M+1},\dotsc,I_{M+L}$.
\begin{enumerate}
    \item WENO-based forecast by 
    \begin{enumerate}
        \item extrapolating the WENO reconstruction: $\b p(t)=\b p^M_\text{WENO}(t)$ for all $t\in \cup_{l=1}^L I_{M+l}$.
        \item taking the $l$-day cell average using the WENO extrapolation: \[\b p(t)\coloneqq \bar{\b p}\coloneqq \frac{1}{l}\b \sum_{m=1}^l\int_{I_{M+m}}p^M_\text{WENO}(\tau)\mathrm d\tau,\quad  t\in \cup_{l=1}^L I_{M+l}.\]
    \end{enumerate}
    \item Taking $\b p(t)\coloneqq\bar{\b p}_M$ from the optimization procedure for all $t\in \cup_{l=1}^L I_{M+l}$.
    \item Taking a historical predictor:
    \begin{itemize}
        \item Compute $\text{argmin}_{m=2,\dotsc,M-1}(\Vert \bar{\b p}_m - \bar{\b p}_M\Vert + \Vert \bar{\b p}_m - \bar{\b p}_{m-1} - (\bar{\b p}_M - \bar{\b p}_{M-1}) \Vert)$, meaning find the index that belongs to the parameter combination which is \emph{the closest neighbor} to the current one in terms of parameter value distance and the historical trend.
        \item Take $\b p(t)\coloneqq\bar{\b p}_{m+1}$ for all $t\in \cup_{l=1}^L I_{M+l}$.
    \end{itemize}
\end{enumerate}
This list of techniques is not complete. For instance, one could include more neighbors into the historical predictor or repeat the procedure for every $l$.

Starting with the right boundary of each cell $I_i$, $i=1,\dotsc,M$, we take the numerical solution obtained by the piecewise constant parameters, see Figure~\ref{fig:PW_and_WENO_sol}, as an initial condition. Then we make predictions using the above techniques for a forecast window of $L\in\{3,5,7,10\}$ days. For each $L$, we plot the results of the top three methods  with respect to  the average cost value. Subsequently, we consider the best and worst prediction of the overall best method, that is, the one with smallest average cost value. In all figures, we state which $k$ is chosen for WENO and emphasize with \enquote{const} or \enquote{lin} the method of constructing the ghost cells.

In addition to plotting the cost reflecting the deviation from the real data, we present two more kinds of plots. The first accounts for the fact, that taking the numerical solution as an initial condition may result already in a large deviation from the real data as the numerical fit is not perfect. This motivates us to shift the initial condition to match the real data and compensate the loss of mass (people) by the compartment $S$. 

The final experiments are concerned with depicting the relative deviation to the numerical solution corresponding to the piecewise constant parameter set -- here the initial data match by construction. 

\begin{figure}[!htbp]
\centering
	\begin{subfigure}[t]{0.495\textwidth}
		\includegraphics[width=\textwidth]{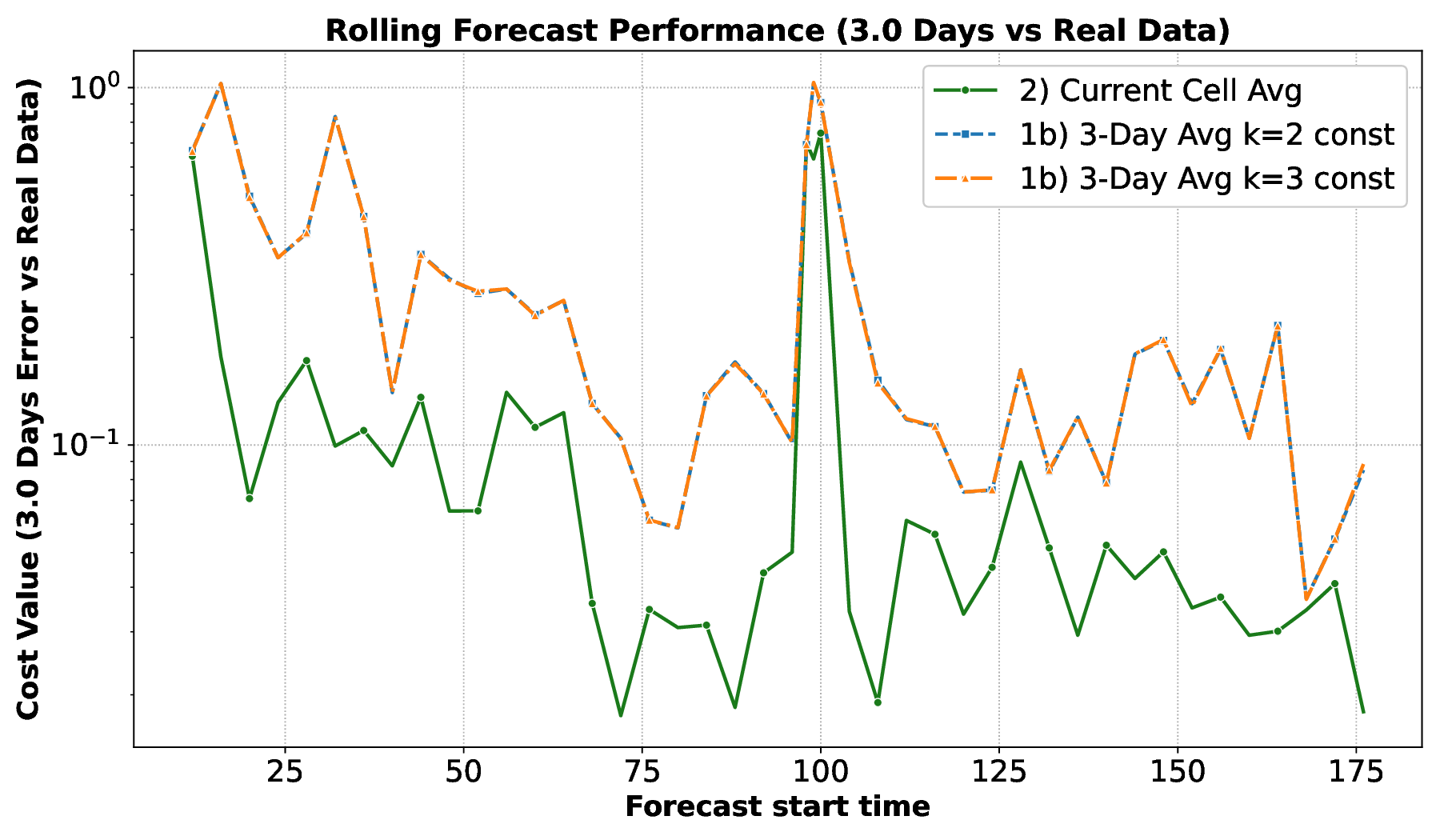}
	\end{subfigure}
	\begin{subfigure}[t]{0.495\textwidth}
		\includegraphics[width=\textwidth]{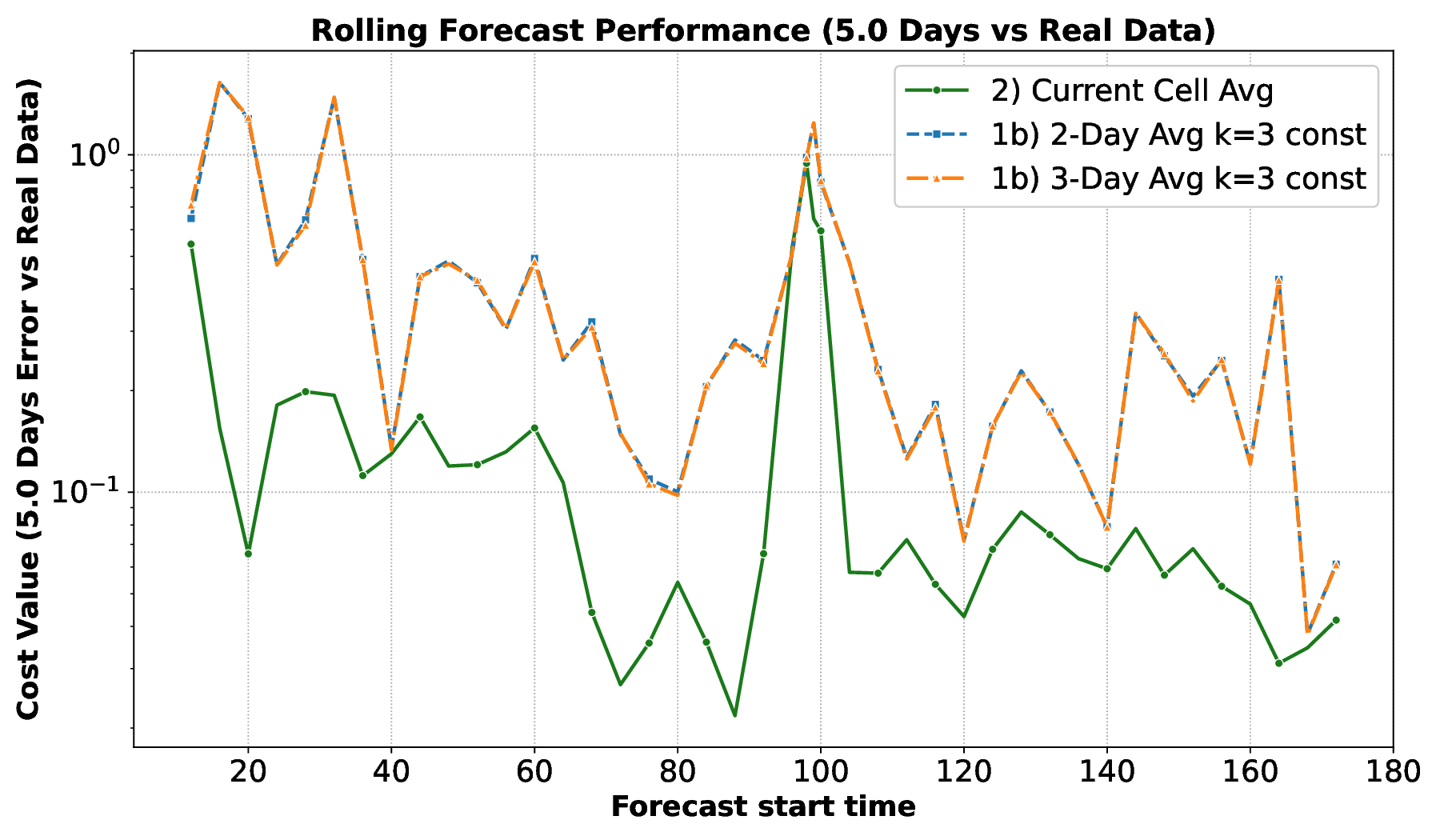}
	\end{subfigure}\\
	\begin{subfigure}[t]{0.495\textwidth}
		\includegraphics[width=\textwidth]{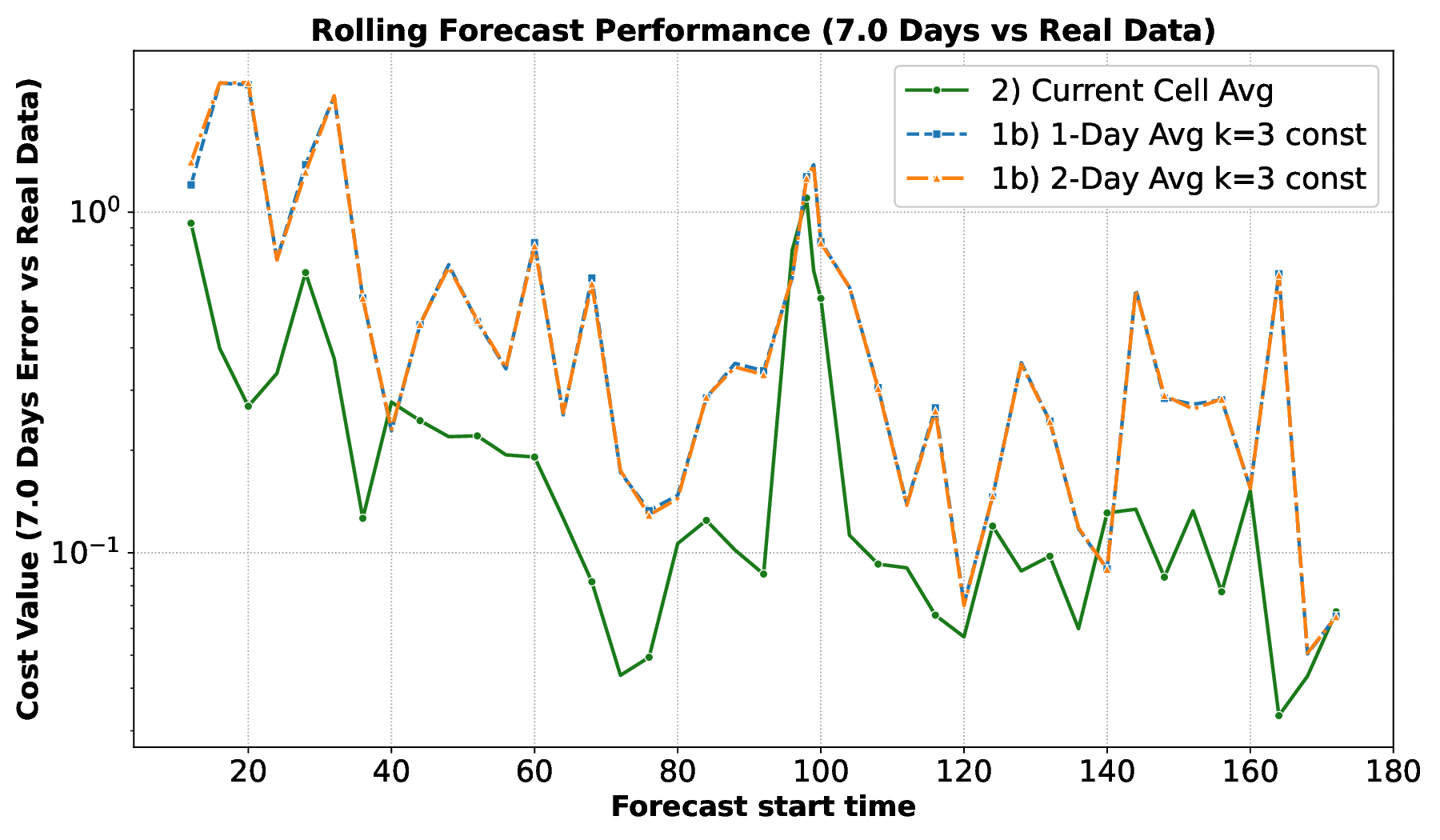}
	\end{subfigure}
    	\begin{subfigure}[t]{0.495\textwidth}
		\includegraphics[width=\textwidth]{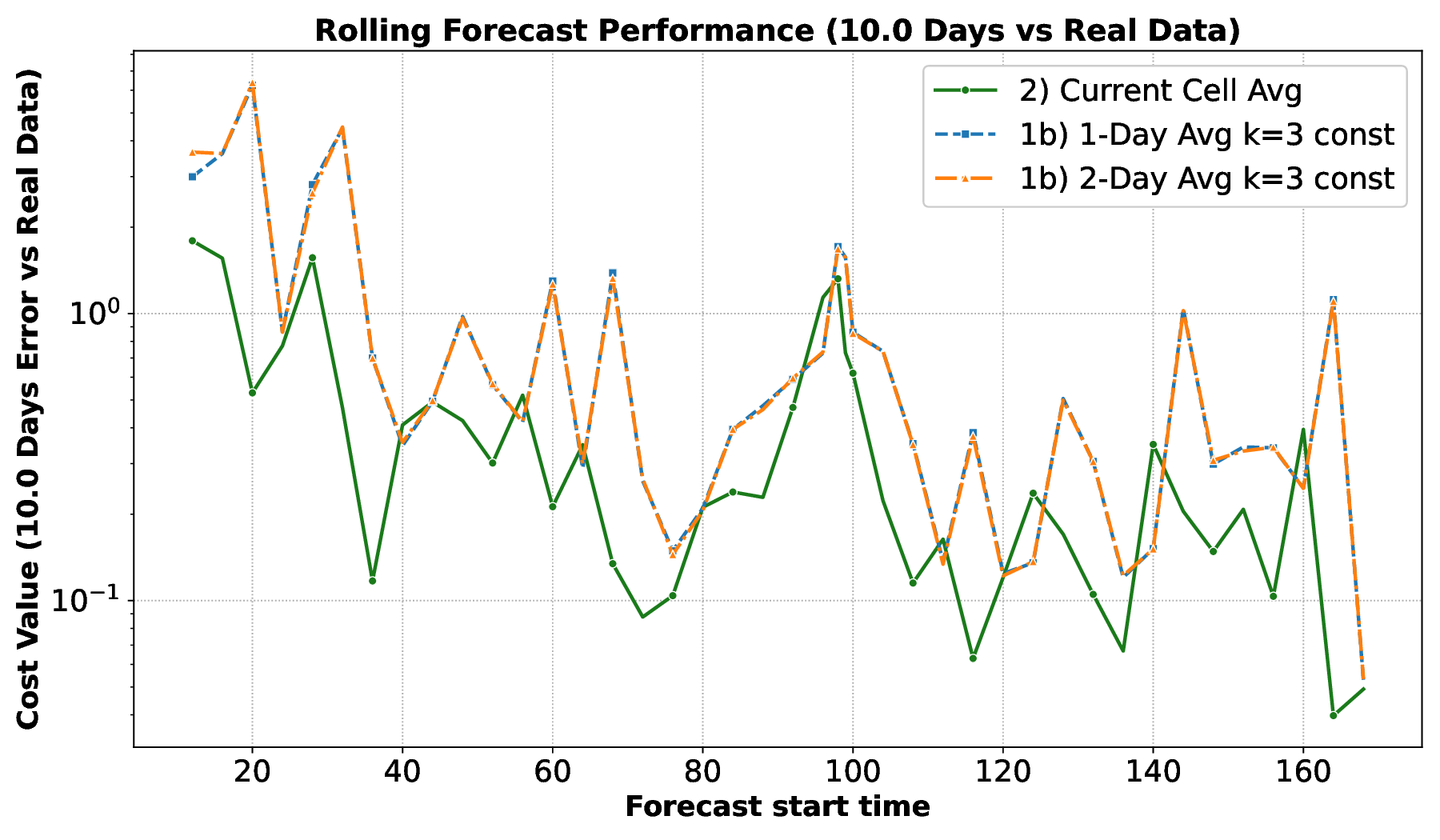}
	\end{subfigure}
	\caption{Costs for various initial conditions and forecast horizons.}\label{fig:forecast_cost}
\end{figure}

\begin{figure}[!htbp]
\centering
	\begin{subfigure}[t]{0.495\textwidth}
		\includegraphics[width=\textwidth]{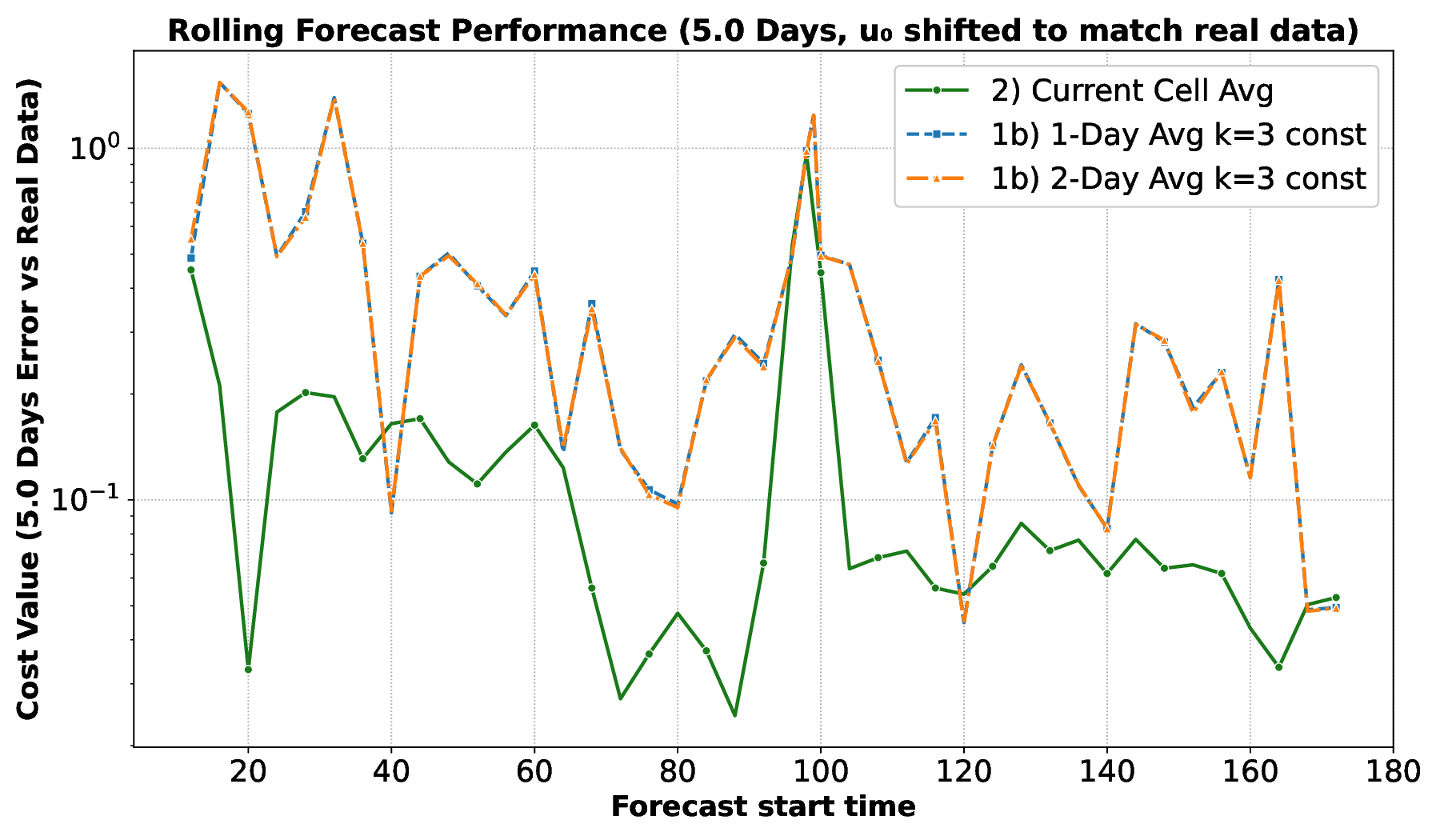}
	\end{subfigure}
	\begin{subfigure}[t]{0.495\textwidth}
		\includegraphics[width=\textwidth]{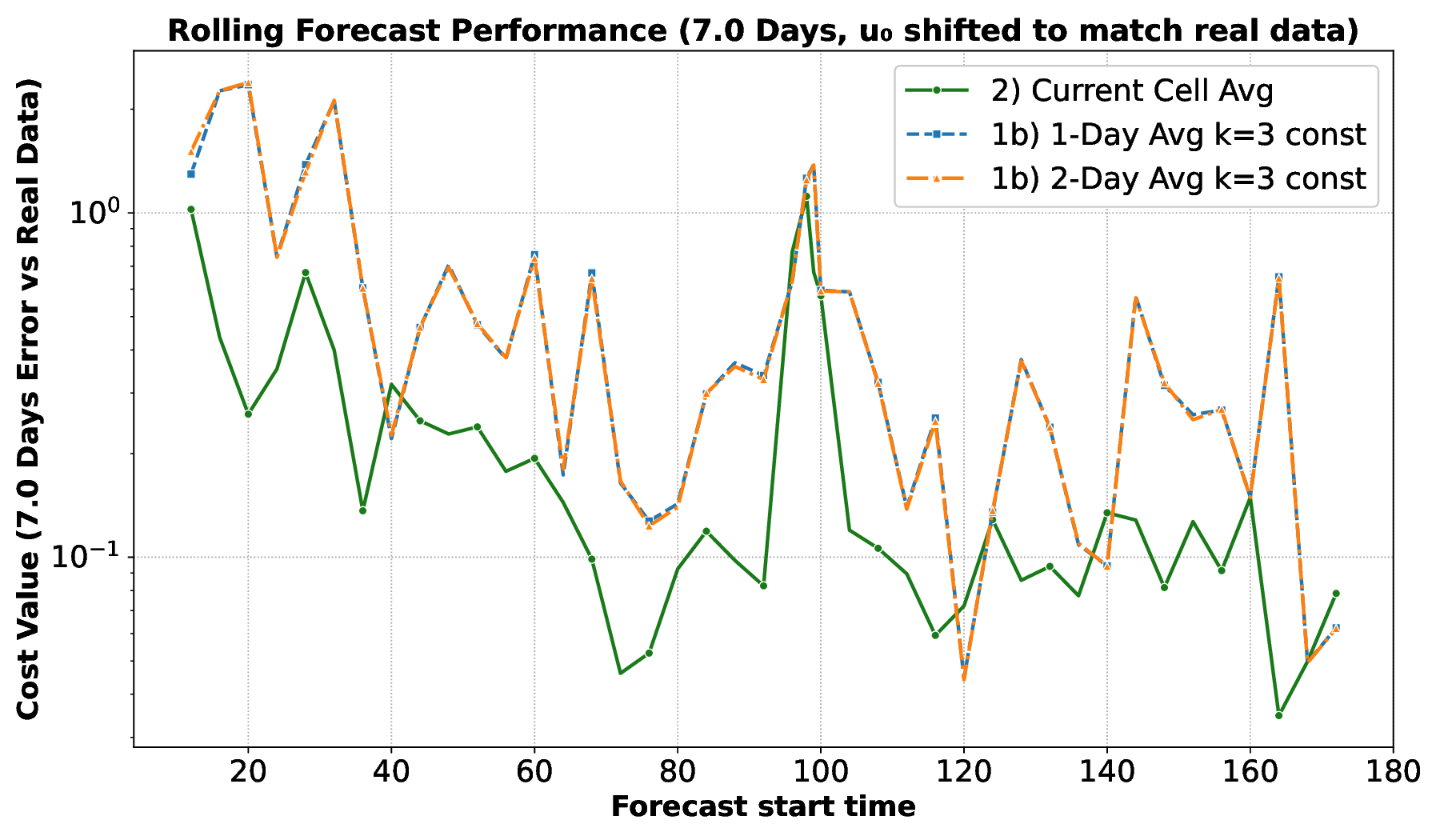}
	\end{subfigure}\\
	\caption{Costs with real-data initial conditions and $L\in\{5,7\}$.}\label{fig:forecast_cost_shifted}
\end{figure}

\begin{figure}[!htbp]
\centering
	\begin{subfigure}[t]{0.495\textwidth}
		\includegraphics[width=\textwidth]{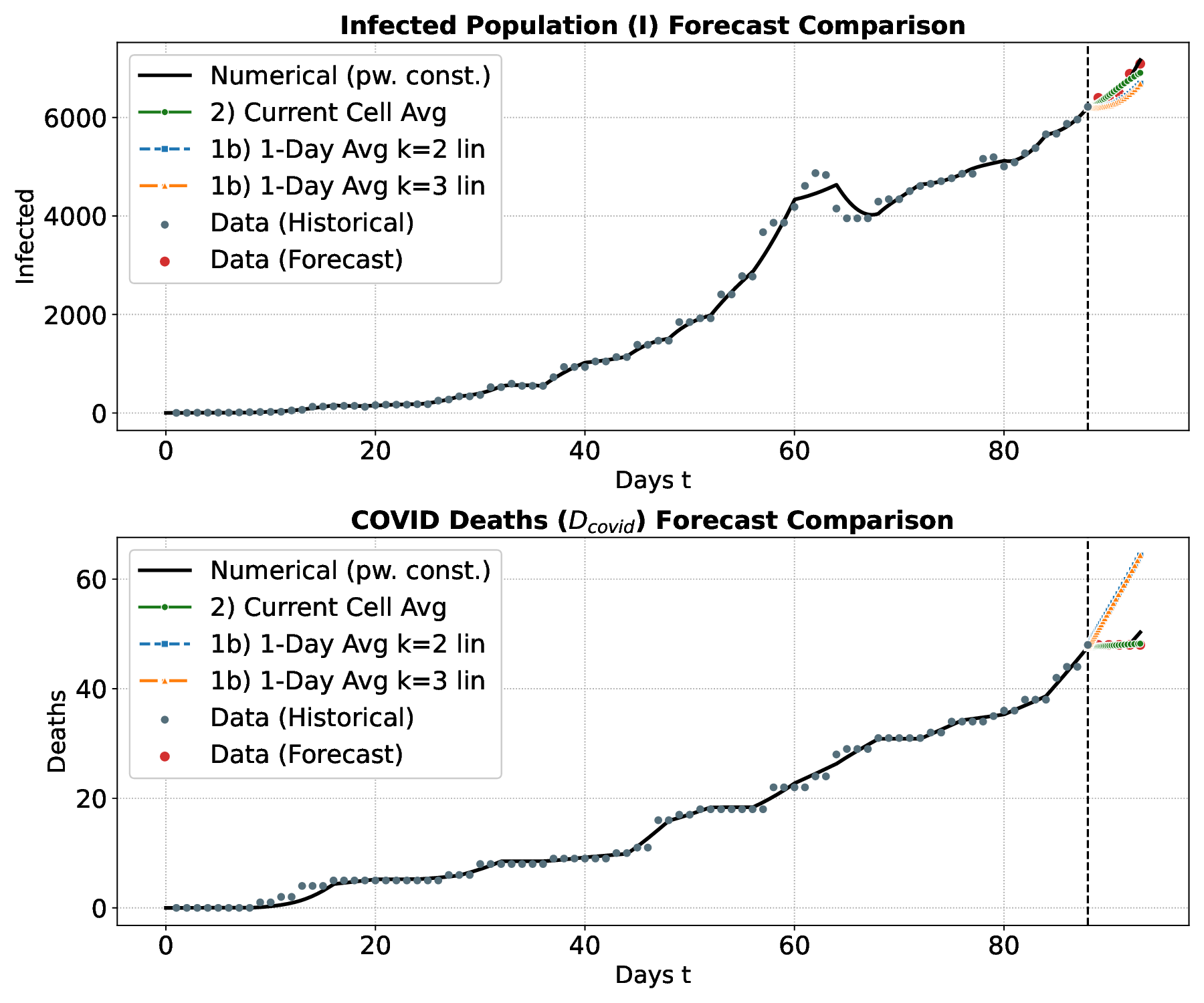}    \subcaption{}
	\end{subfigure}
	\begin{subfigure}[t]{0.495\textwidth}
		\includegraphics[width=\textwidth]{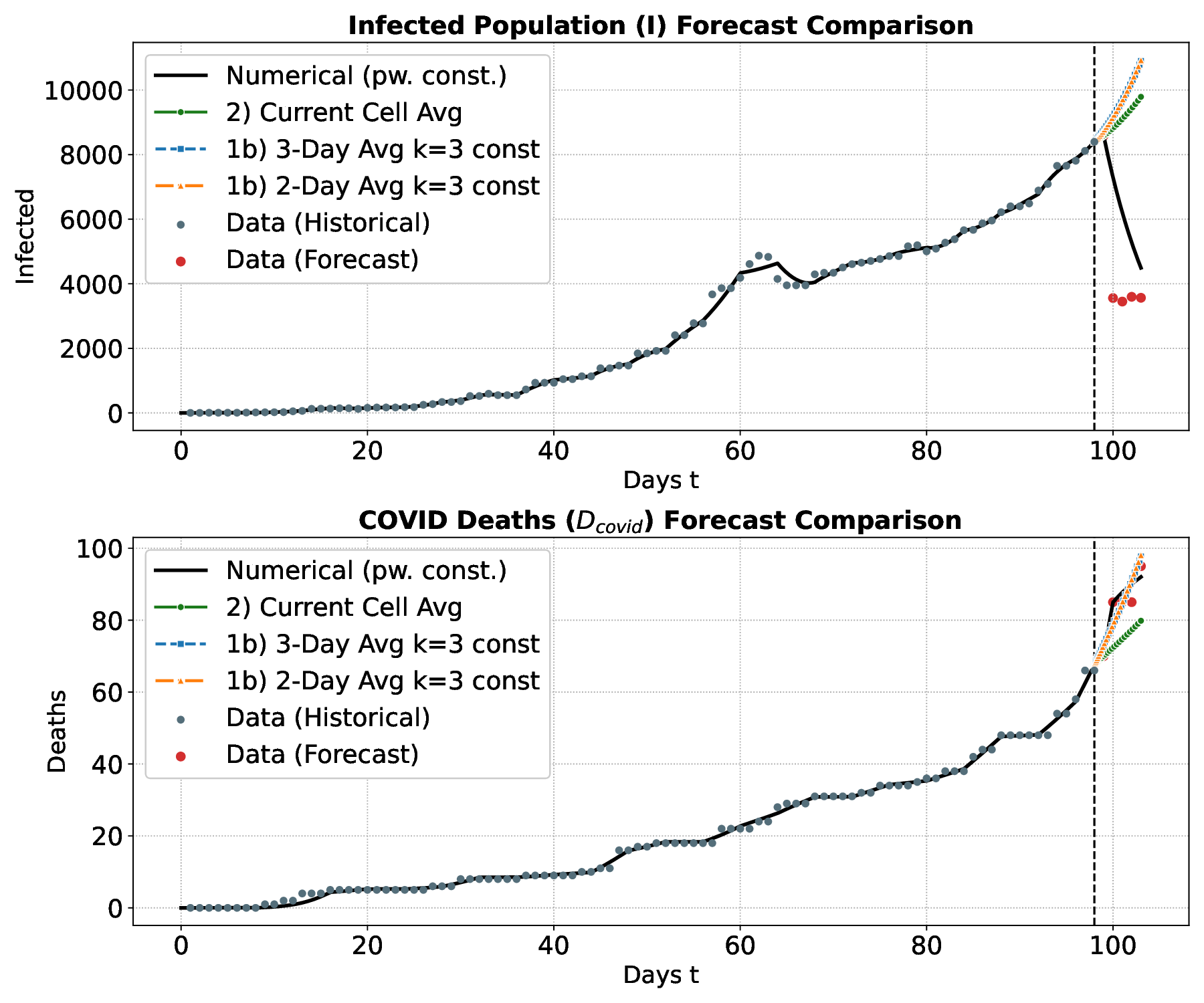}    \subcaption{}
	\end{subfigure}
    \\
    	\begin{subfigure}[t]{0.495\textwidth}
		\includegraphics[width=\textwidth]{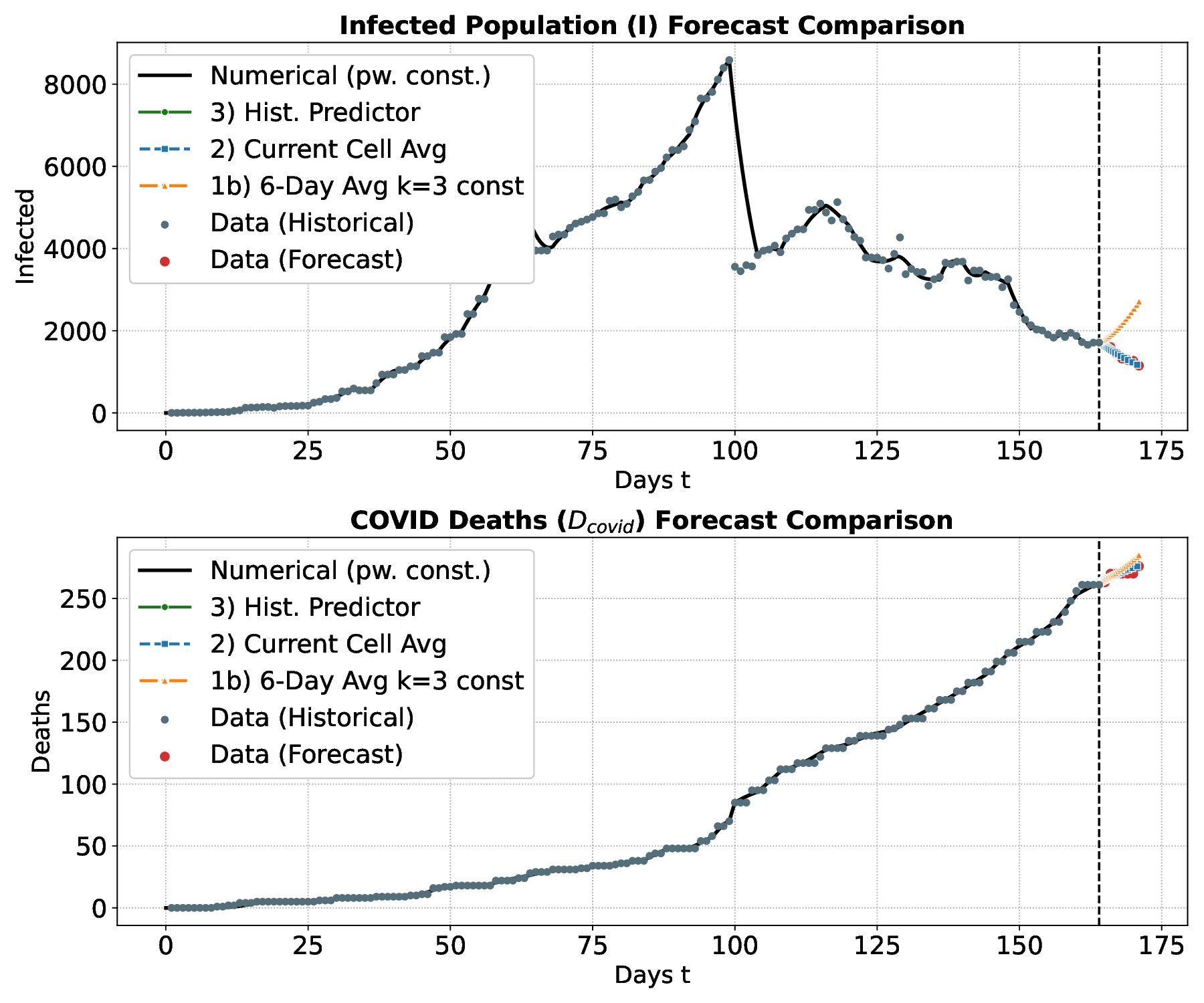}    \subcaption{}
	\end{subfigure}
	\begin{subfigure}[t]{0.495\textwidth}
		\includegraphics[width=\textwidth]{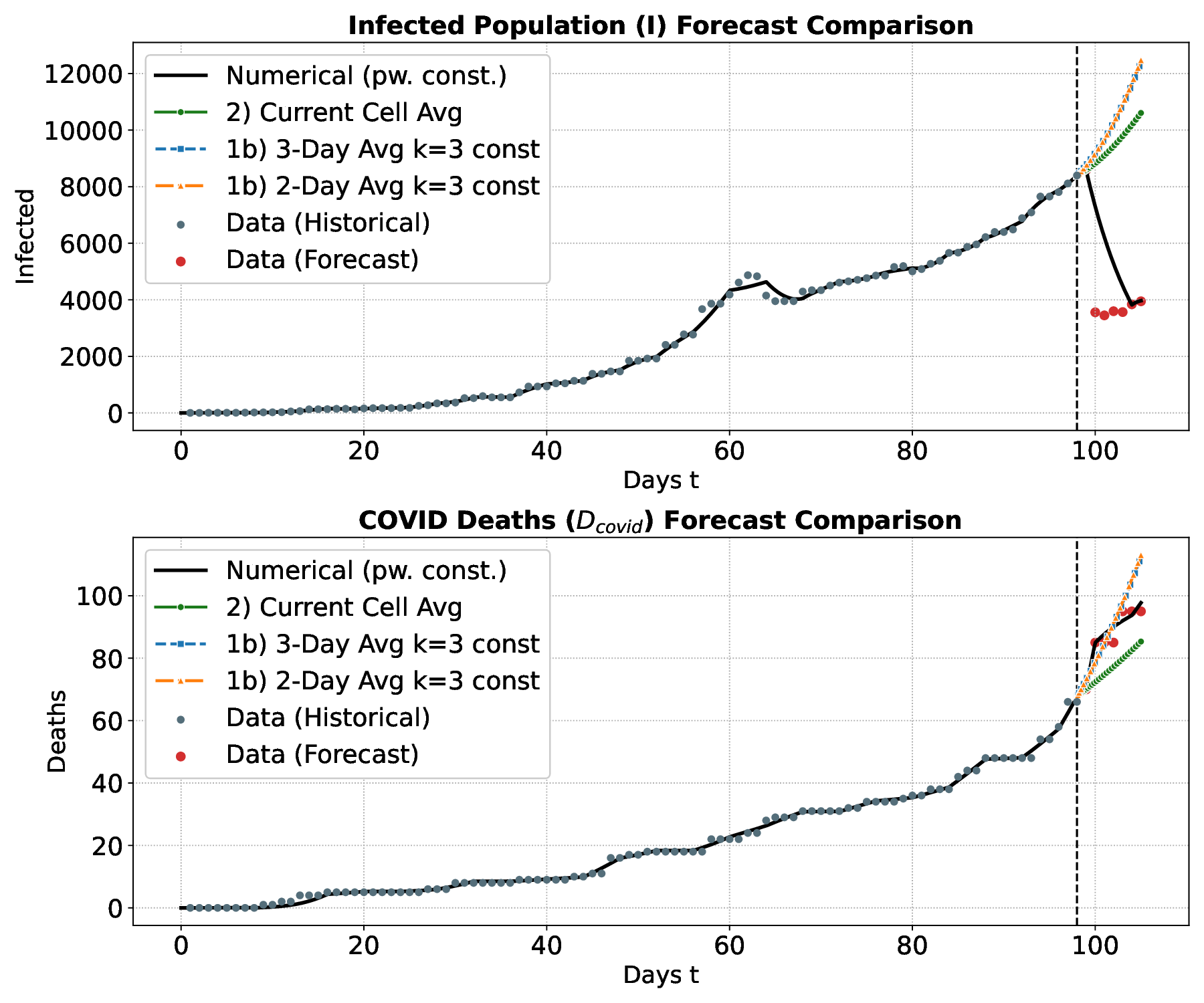}    \subcaption{}
	\end{subfigure}\\
      	\begin{subfigure}[t]{0.495\textwidth}
		\includegraphics[width=\textwidth]{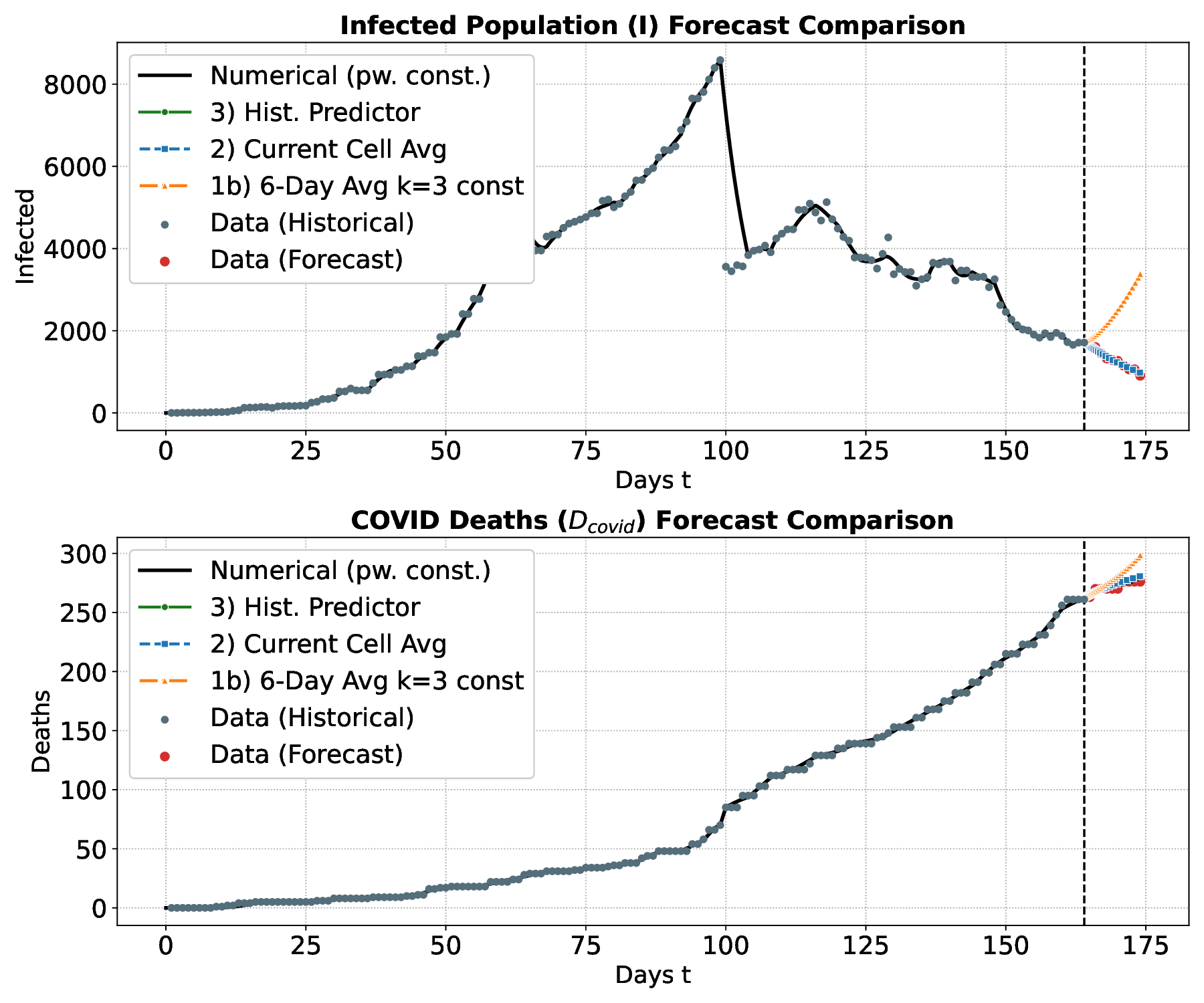}    \subcaption{}
	\end{subfigure}
	\begin{subfigure}[t]{0.495\textwidth}
		\includegraphics[width=\textwidth]{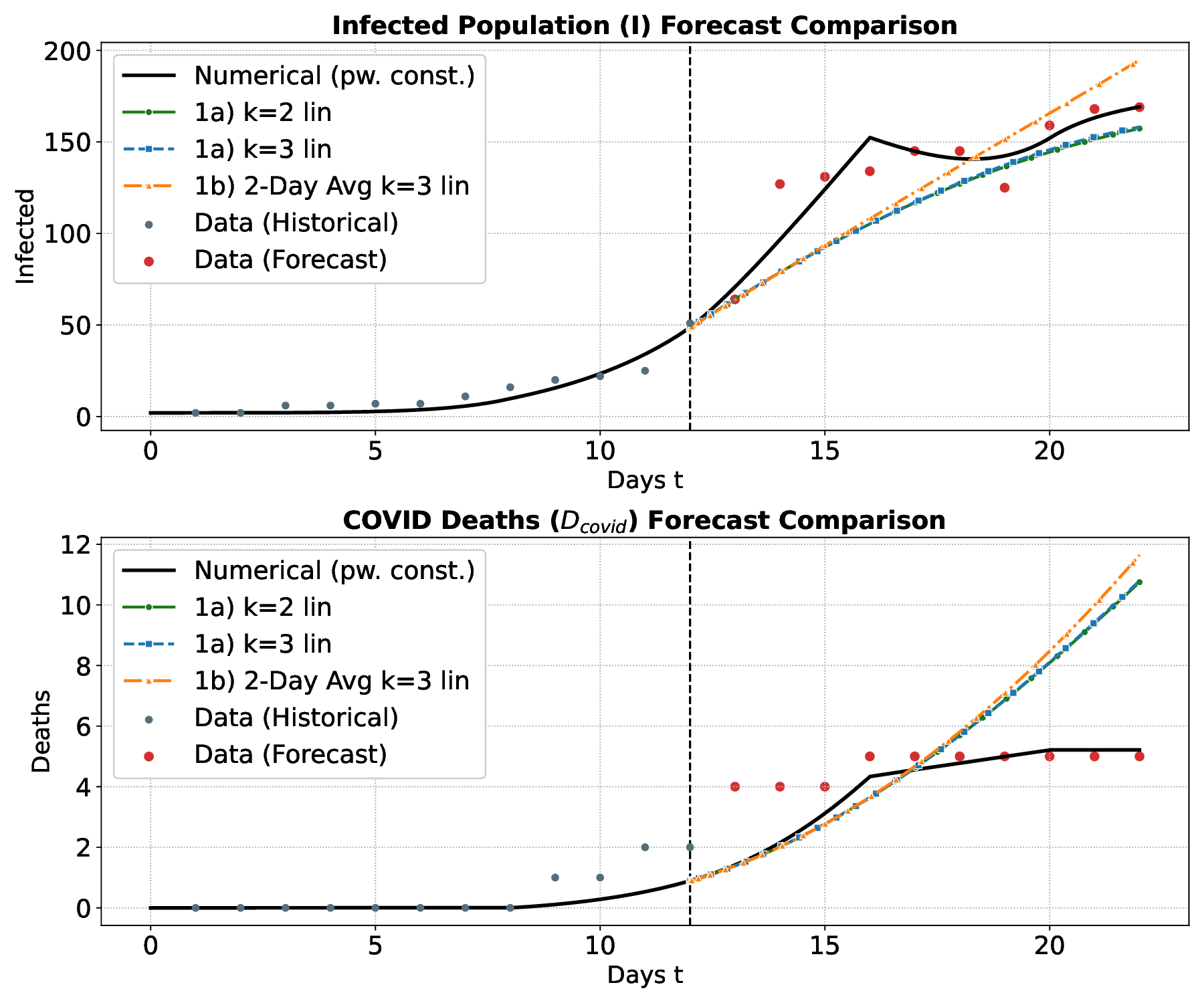}    \subcaption{}
	\end{subfigure}
	\caption{Prediction of $L\in\{5,7,10\}$ days (from top to bottom). Left: Best prediction. Right: worst prediction of best predictor and the corresponding forecasts of other predictors.}\label{fig:best_worst}
\end{figure}

While we present the plots of the cost function for all $L$, we restrict to fewer plots when it comes to the comparison to the numerical solution and the comparison of the shifted numerical solution to real data in order to maintain readability. Nevertheless, all figures will be available in our reproducibility repository \cite{AIM2026repository}, where the interested reader can also run the julia code with a single command reproducing all figures of this work.
The results with the cost values are shown in Figure~\ref{fig:forecast_cost}. As one can see, taking the current cell average as a predictor is the best and achieves around the same level of accuracy using a 3-day and 5-day forecast. In particular the average deviation from the real data is around 10\%. 
As one can see in Figure~\ref{fig:forecast_cost_shifted}, the prediction is even better if real data is used within the initial condition -- which constitutes the more realistic use case anyway.
To illustrate how good or bad the prediction can be, we plot best and worst approximation of the overall best prediction method in Figure~\ref{fig:best_worst}. 

\begin{figure}[!htbp]
\centering
	\begin{subfigure}[t]{0.495\textwidth}
		\includegraphics[width=\textwidth]{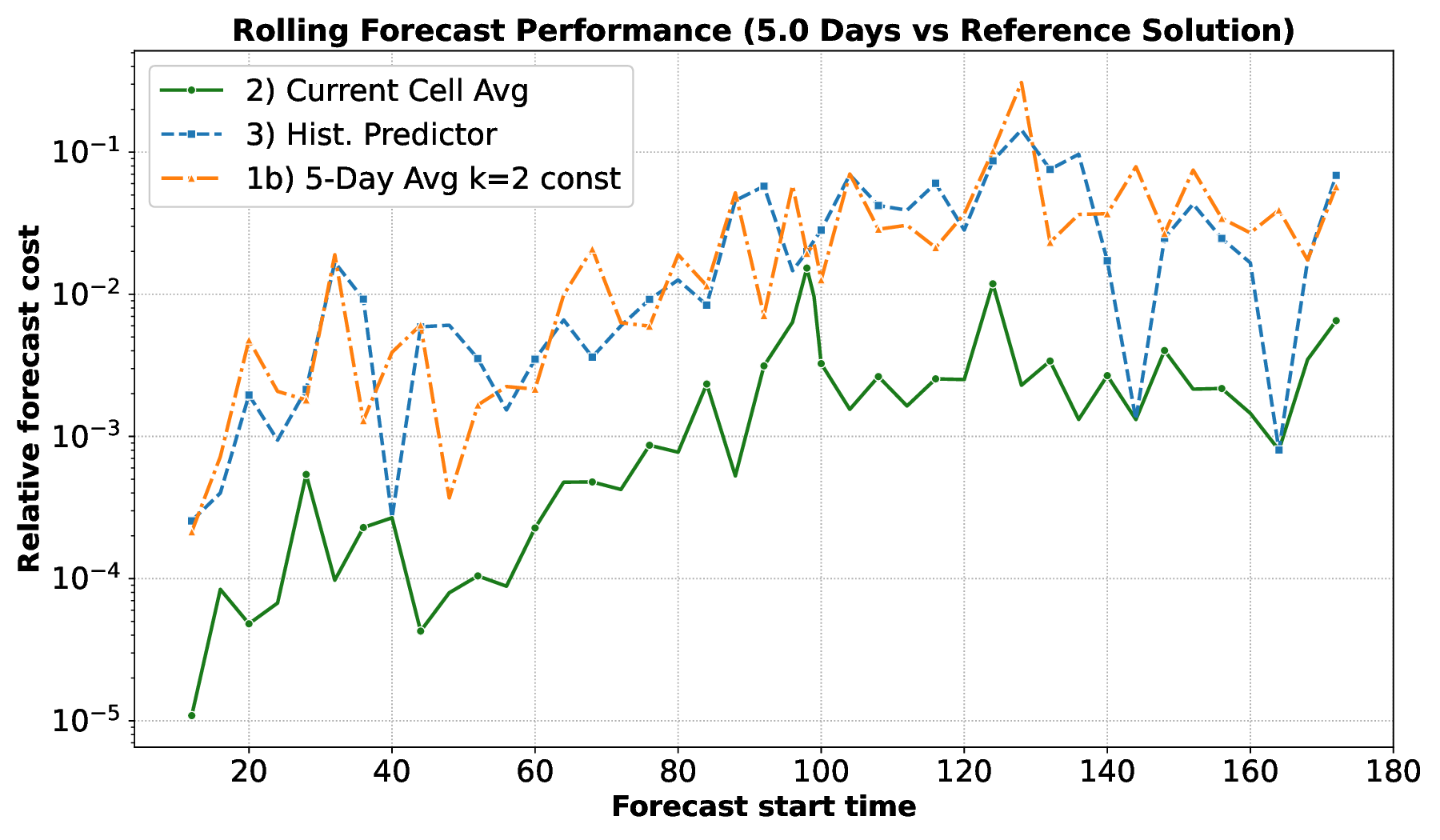}
	\end{subfigure}
	\begin{subfigure}[t]{0.495\textwidth}
		\includegraphics[width=\textwidth]{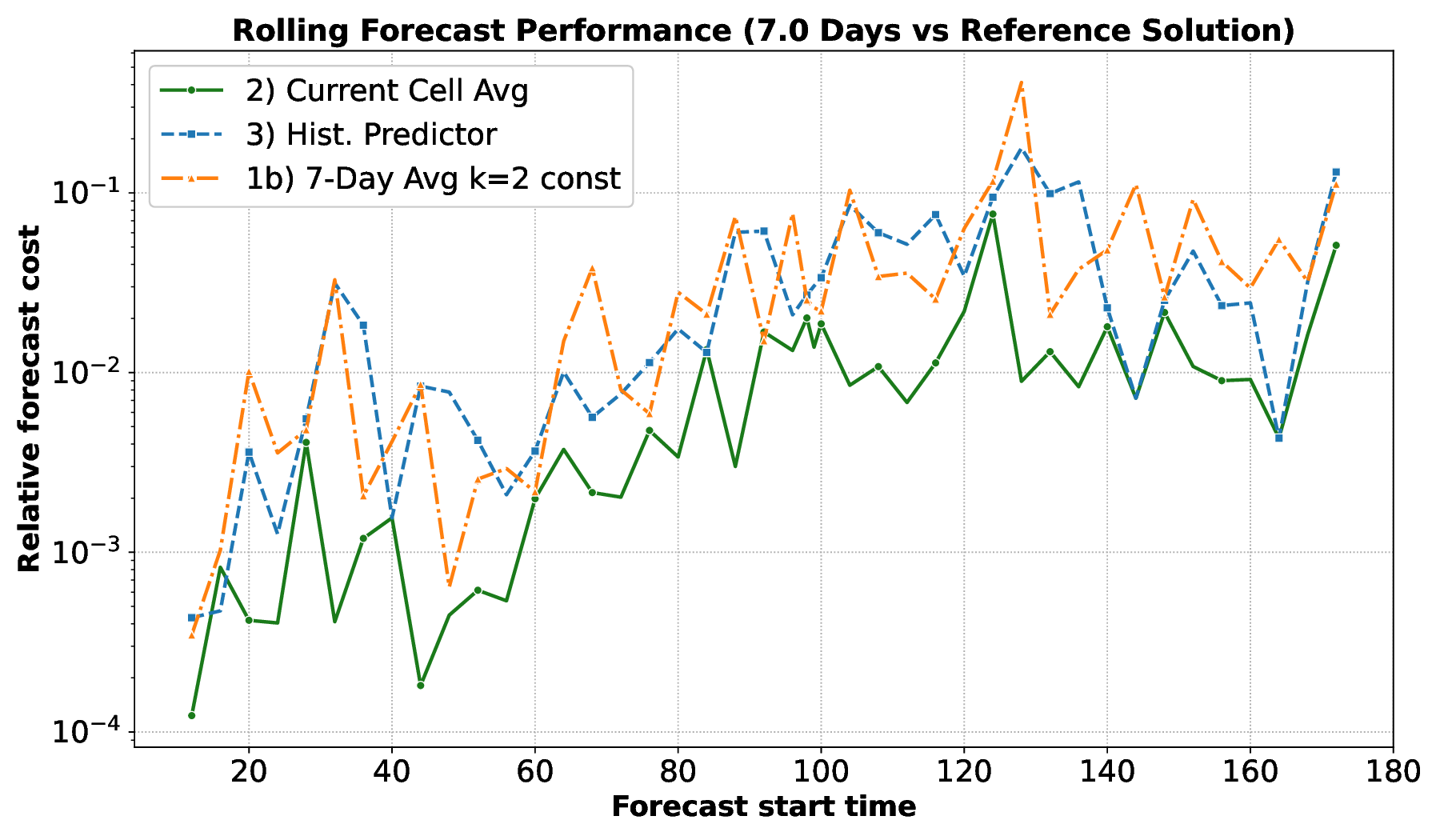}
	\end{subfigure}\\
	\caption{Relative error of prediction and numerical solution $\b u$ for $L\in\{5,7\}$.}\label{fig:forecast_rel_err_to_num}
\end{figure}

Finally, looking on the relative error to the numerical solution, depicted in Figure~\ref{fig:forecast_rel_err_to_num}, we observe a small relative error at smaller times while the error grows at later points in time. This is no surprise as the dynamics at the beginning are slowly changing and hence, taking old parameters may be good enough. Still, as can be observed in Figure~\ref{fig:PW_and_WENO_sol}, the optimization procedure did not use parameters that are relatively close to each other, which is why we may use a different tool in future works.     In any case, we want to emphasize that the used data is only a lower bound for the true figures, so that the WENO approach may be interpreted as a worst-case scenario, see Figure~\ref{fig:PW_and_WENO_sol}.
\subsection{Notes on the Implementation}
Our code \cite{AIM2026repository} is such that one can replace the data by a data base for any Country and disease. Although we assumed some parameters to be constant, this can be reversed by simply commenting out a few lines of code. Additionally, one can choose the parameters to be optimized, which may be interesting, if one uses a sub-model of \eqref{eq:compr_model}. Finally, the forecast functions can easily be adapted to take also data for further compartments into account by following the existing structure -- the optimization tool already does include, e.g. vaccinations or hospitalizations, which however were not used in this proof-of-concept study.




\section{Summary and Outlook}\label{sec:summary}
In this study, we presented a comprehensive system of ODEs to model a pandemic or epidemic evolution and demonstrated how to estimate solution- and time-dependent coefficients. To that end, we applied  unconditionally positive and conservative MPRK schemes to preserve the model's properties on a numerical basis. We then used the numerical approximation within a cost function, whose minima was search by means of machine learning, yielding a vector of piecewise constant parameters. In a post-processing procedure, we reconstructed the time-dependent coefficients by means of a WENO reconstruction. 

We demonstrated that our approach is capable of reproducing the COVID-19 dynamics in Ghana, as well as making predictions that are only approximately 10\% off the real data and computed by solving a system of ODEs numerically. This is beyond the state-of-the-art since with our technique we are optimizing only the system of ODEs to better represent the data and then solve it with high-order schemes, meaning improving the numerical method will provably improve the numerical approximations and predictions. 

Future research will focus on refining the optimization procedure to better tailor the underlying systems of differential equations to the specific dynamics of a given disease. Additionally, we plan to evaluate our methodology across a broader suite of frameworks—specifically Models 2–5 outlined in Table~\ref{tab:model_para}—utilizing empirical epidemiological data from Germany. To support this expanded scope of applications, the codebase will be systematically upgraded to maximize portability and user accessibility. Finally, subsequent iterations of this work will explicitly account for non-pharmaceutical interventions and political actions by incorporating mechanisms to suppress or dynamically estimate target transmission rates prior to forecasting.

\section*{Acknowledgments}

T.\ Izgin thanks Chi-Wang Shu from Brown University for many fruitful discussions, and Janina Bender from the University of Kassel for developing the script bayesian\_tools.jl in \cite{AIM2026repository}. 
The authors also acknowledge the use of Google Antigravity IDE (Individual Preview Plan) for autonomous code generation and workspace management. Specifically, parallel sub-agents were utilized to draft initial data visualization scripts. All AI-generated code and logic were subsequently audited, verified, and manually refined by the authors to ensure scientific accuracy.

\section*{Declarations}

\bmhead{Funding}
The author A. Meister gratefully acknowledges the financial support by the Deutsche Forschungsgemeinschaft (DFG) through the grant ME 1889/12-1 (project number 56008780).
The author T. Izgin acknowledges the financial support by Fulbright Germany.

\bmhead{Conflict of interest}
The authors declare that they have no conflict of interest.

\bmhead{Availability of code, data, and materials}
The source code is available via the repository \cite{AIM2026repository}.

\bmhead{Authors' contributions}

Conceptualization: Thomas Izgin, Andreas Meister, Isaac Azure;

Data curation: Thomas Izgin;

Formal analysis and investigation: Thomas Izgin, Andreas Meister, Isaac Azure;

Methodology: Thomas Izgin, Andreas Meister;

Project administration: Thomas Izgin, Andreas Meister;

Software: Thomas Izgin, Janina Bender;

Visualization: Thomas Izgin;

Writing - original draft preparation: Thomas Izgin, Isaac Azure;

Writing - review and editing: Thomas Izgin, Andreas Meister, Isaac Azure.

\appendix

\renewcommand{\arraystretch}{1.3}

\renewcommand{\a}[2]{a^{#1}_{#2}}
\renewcommand{\r}[2]{r^{#1}_{#2}}
\renewcommand{\arraystretch}{1.1} 

\newgeometry{margin = 5mm}
\section{Model Parameters from the Literature}\label{app:model_para}
\begin{table}[!htbp]
	\centering
	\caption{Description of Parameter Notations of the Model}
	\begin{tabular}{lp{11cm}}
		\hline
		\textbf{Parameter} & \textbf{Description} \\
		\hline
		$\delta_{V} \in \{0,1\}$ & Modification parameter of death in vaccinated individuals. \\ 
		$\delta_{E}\in \{0,1\}$ & Modification parameter of death in exposed class. \\ 
		$\delta_{L}\in \{0,1\}$ & Modification parameter of death in latent class. \\ 
		$\delta_{H}\in \{0,1\}$ & Modification parameter of death in hospitalized individuals. \\ 
		$\delta_{Q}\in \{0,1\}$ & Modification parameter of death in quarantined individuals. \\ 
		$p$ & Proportion of recruited individuals who are vaccinated. \\ 
		$\Lambda$ & Recruitment or birth rate into the population. \\ 
		$\mu$ & Natural death rate. \\ 
		$\phi$ & Fraction of exposed individuals progressing to infection. \\ 
		$\psi$ & Fraction of infected individuals recovering without hospitalization. \\ 
		$\gamma$ & Fraction of latent individuals progressing to infection. \\ 
		$\mu_{V_E}$ & Decay rate of the virus in the environment. \\ 
		$\alpha_I$ & Disease-induced death rate for infectious individuals. \\ 
		$\alpha_H$ & Disease-induced death rate for hospitalized individuals. \\ 
		$a_S^E$ & Transmission rate from susceptible individuals to exposed due to contact with infectious or latent individuals. \\ 
		$a_S^V$ & Rate at which susceptible individuals get vaccinated. \\ 
		$a_V^S$ & Rate of waning vaccine immunity returning vaccinated individuals to susceptible class. \\ 
		$a_V^E$ & Infection rate of vaccinated individuals. \\ 
		$a_E^I$ & Rate at which exposed individuals become infectious. \\ 
		$a_E^S$ & Possible reverse movement from exposed to susceptible. \\ 
		$a_E^Q$ & Rate at which exposed individuals are quarantined. \\ 
		$a_E^L$ & Rate at which exposed individuals enter latent stage. \\ 
		$a_L^I$ & Rate at which latent individuals become infectious. \\ 
		$a_L^R$ & Recovery rate of latent individuals. \\ 
		$a_L^Q$ & Quarantine rate of latent individuals. \\ 
		$r_L^{V_E}$ & Shedding rate of virus from latent individuals into the environment. \\ 
		$a_I^R$ & Recovery rate of infectious individuals. \\ 
		$a_I^Q$ & Quarantine rate of infectious individuals. \\ 
		$a_I^H$ & Hospitalization rate of infectious individuals. \\ 
		$r_I^{V_E}$ & Shedding rate of virus from infectious individuals into the environment. \\ 
		$a_H^R$ & Recovery rate of hospitalized individuals. \\ 
		$a_R^S$ & Rate of loss of immunity returning recovered individuals to susceptible class. \\ 
		$a_Q^R$ & Recovery rate of quarantined individuals. \\
		\hline
	\end{tabular}
	\label{tab:parameters}
\end{table}

\begin{table}[h!]
	\centering
	\renewcommand{\arraystretch}{1.3}
	\setlength{\tabcolsep}{12pt}
	\caption{Initial Conditions of State Variables for Five Models}\label{tab:IC}
	\begin{tabular}{lccccc}
		\toprule
		\textbf{Variable} & \textbf{Model 1} & \textbf{Model 2} & \textbf{Model 3} & \textbf{Model 4} & \textbf{Model 5} \\
		\midrule
		$S(0)$ & 30416000 & 2500 & 338000000 & 338173377 & 1000000 \\ 
		$V(0)$ & 0 & 10 & 0 & 10000 & 35 \\ 
		$E(0)$ & 5 & 20 & 50000 & 10000 & 50 \\ 
		$L(0)$ & 5 & 3 & 0 & 0 & 0 \\ 
		$I(0)$ & 2 & 70 & 30000 & 56029 & 32 \\ 
		$H(0)$ & 0 & 3 & 100000 & 40450 & 0 \\ 
		$R(0)$ & 0 & 1 & 10000 & 10000 & 15 \\ 
		$Q(0)$ & 0 & 0 & 0 & 0 & 10 \\ 
		$D(0)$ & 0 & 0 & 0 & 0 & 0 \\ 
		Reference   & \cite{moore2022global}              & \cite{diagne2021mathematical}     & \cite{santosh2025novel}  & \cite{rattanakul2024mathematical} & \cite{haq2022new} \\
		\bottomrule
	\end{tabular}
\end{table}

\begin{landscape}

	\begin{center}
		\footnotesize 
		\setlength{\tabcolsep}{4pt} 
		\begin{longtable}{cccccc}
			\caption{Model parameters}\label{tab:model_para}\\
			\toprule
			\textbf{Parameters} & \textbf{Model 1} & \textbf{Model 2} & \textbf{Model 3} & \textbf{Model 4} & \textbf{Model 5} \\
			\midrule
			\endfirsthead
			
			\multicolumn{6}{c}{{\tablename\ \thetable{} }} \\
			\toprule
			\textbf{Parameters} & \textbf{Model 1} & \textbf{Model 2} & \textbf{Model 3} & \textbf{Model 4} & \textbf{Model 5} \\
			\midrule
			\endhead
			
			\midrule
			\multicolumn{6}{r}{{Continued on next page}} \\
			\bottomrule
			\endfoot
			
			\bottomrule
			\endlastfoot
			
			$\delta_{V}$ & 0 & 1 & 0 & 0 & 1\\
			$\delta_{E}$ & 1 & 1 & 0 & 1 & 1\\
			$\delta_{L}$ & 1 & 1 & 0 & 0 & 0\\
			$\delta_{H}$ & 0 & 1 & 0 & 1 & 0\\
			$\delta_{Q}$ & 0 & 0 & 0 & 0 & 1\\
			$p$ & 0 & 0.0001 & 0 & 0 & $1/40$\\
			$\Lambda$ & $1319.294$ & $\dfrac{10000}{59\times365}$ & 0 & 3251 & $0.46$\\
			$\mu$ & $0.000042578$ & $\dfrac{1}{59\times365}$ & 0 & $0.3349\times10^{-4}$ & $0.0991$\\
			$\phi$ & $0.01000001$ & 0.7 & 1 & 1 & 1\\
			$\psi$ & 1 & $0.05$ & $1/2$ & $1/2$ & 0\\
			$\gamma$ & $0.00500005$ & 0.86 & 0 & 0 & 0\\
			$\mu_{V_E}$ & $0.290000$ & 0 & 0 & 0 & 0\\
			$\alpha_I$ & $0.00440000$ & $0.018$ & 0 & $1/16.1$ & 0\\
			$\alpha_H$ & 0 & $0.018$ & 0 & $1/11.2$ & 0\\
			$\a{S}{E}$ & $6.038\times10^{-8}(I + 0.62811041\,L) - 4.00199\times10^{-8}V_E$ & $\frac{1.12(0.3L + 1.8I + 0.3H)}{S+V+E+I+L+H+R}$ & $0.3 I$ & $9.1508\times10^{-9} I$ & $0.999(S+V+E+I+Q+R)(E+I)$\\
			$\a{S}{V}$ & 0 & 0.4 & 0 & 0 & 0.4\\
			$\a{V}{S}$ & 0 & 0 & 0 & 0 & 0\\
			$\a{V}{E}$ & 0 & $0.2\a{S}{E}$ & 0 & 0 & 0.3002\\
			$\a{E}{I}$ & 0.07142 & 0.13 & 0.1 & 1/4.2 & 0.001\\
			$\a{E}{S}$ & 0 & 0 & 0 & 0 & 0\\
			$\a{E}{Q}$ & 0 & 0 & 0 & 0 & 0.2\\
			$\a{E}{L}$ & 0.14285824 & 0.13 & 0 & 0 & 0\\
			$\a{L}{I}$ & 0.20005051 & 0.13978 & 0 & 0 & 0\\
			$\a{L}{R}$ & 0.79999398 & 0.13978 & 0 & 0 & 0\\
			$\a{L}{Q}$ & 0 & 0 & 0 & 0 & 0\\
			$\r{L}{V_E}$ & 0.01780400 & 0 & 0 & 0 & 0\\
			$\a{I}{R}$ & 0.0805840 & 0.86 & $0.1$ & $2/30$ & 0\\
			$\a{I}{Q}$ & 0 & 0 & 0 & 0 & 0.005\\
			$\a{I}{H}$ & 0 & 0.86 & $0.04$ & $2/1.5$ & 0\\
			$\r{I}{V_E}$ & 0.92152716 & 0 & 0 & 0 & 0\\
			$\a{H}{R}$ & 0 & 0.0701 & 0.03 & $1/11.5$ & 0\\
			$\a{R}{S}$ & 0.41138431 & 0.011 & 0 & 0.0426 & 0\\
			$\a{Q}{R}$ & 0 & 0 & 0 & 0 & $1/14$\\
			Reference   & \cite{moore2022global}              & \cite{diagne2021mathematical}     & \cite{santosh2025novel}  & \cite{rattanakul2024mathematical} & \cite{haq2022new} \\
			
		\end{longtable}
	\end{center}
	
\end{landscape}\restoregeometry

\bibliography{sn-bibliography}

\end{document}